\documentclass[12pt]{article}
\pdfoutput=1
\usepackage[]{graphicx}
\usepackage[]{color}

\makeatletter
\def\maxwidth{ %
  \ifdim\Gin@nat@width>\linewidth
    \linewidth
  \else
    \Gin@nat@width
  \fi
}
\makeatother

\definecolor{fgcolor}{rgb}{0.345, 0.345, 0.345}

\usepackage{framed}
\makeatletter
 {\par\unskip\endMakeFramed%
 \at@end@of@kframe}
\makeatother

\definecolor{shadecolor}{rgb}{.97, .97, .97}
\definecolor{messagecolor}{rgb}{0, 0, 0}
\definecolor{warningcolor}{rgb}{1, 0, 1}
\definecolor{errorcolor}{rgb}{1, 0, 0}
\newenvironment{knitrout}{}{} % an empty environment to be redefined in TeX

\usepackage{alltt}
\usepackage{amsmath}
\usepackage{graphicx}
\usepackage{enumerate}
\usepackage{natbib}
\usepackage{url} % not crucial - just used below for the URL
\usepackage{floatpag}\floatpagestyle{empty}
\usepackage{etoolbox}
\newtoggle{nodata} %% Suppress printing the count data from NATSAL, GUMCAD, GUM Anon and GMSHS (keeping the denominators) 
\toggletrue{nodata}
\usepackage{tikz}

\usetikzlibrary{calc,fit,positioning,arrows,shapes,backgrounds}
\tikzset{>={latex}}
\definecolor{dagblue}{rgb}{0.81,0.902,0.957}
\usepackage{bm}
\graphicspath{{/home/chris/uncertainty/evi/hiv/},{c:/Users/chris/Dropbox/work/uncertainty/evi/hiv/}}

\DeclareMathOperator*{\argmin}{arg\,min}

\newcommand{\blind}{1}

\newcommand{\cov}{\mbox{cov}}
\newcommand{\var}{\mbox{var}}
\newcommand{\logit}{\mbox{logit}}
\newcommand{\y}{\mathbf{y}}
\newcommand{\x}{\mathbf{x}}
\newcommand{\cc}{\mathbf{c}}
\newcommand{\pinodelta}{\overline{(\pi\delta)}}
\newcommand{\pidelta}{(\pi\delta)}
\IfFileExists{upquote.sty}{\usepackage{upquote}}{}
\IfFileExists{upquote.sty}{\usepackage{upquote}}{}
\IfFileExists{upquote.sty}{\usepackage{upquote}}{}
\begin{document}

\def\spacingset#1{\renewcommand{\baselinestretch}%
{#1}\small\normalsize} \spacingset{1}

%%%%%%%%%%%%%%%%%%%%%%%%%%%%%%%%%%%%%%%%%%%%%%%%%%%%%%%%%%%%%%%%%%%%%%%%%%%%%%

\if1\blind
{
  \title{\bf Value of Information: Sensitivity Analysis and Research Design in Bayesian Evidence Synthesis}
  \author{Christopher Jackson, Anne Presanis, Stefano Conti, Daniela De Angelis\thanks{
    This work was funded by the Medical Research Council, grant code U105260566}\hspace{.2cm}\\
    MRC Biostatistics Unit, University of Cambridge; NHS England\\
  }
  \maketitle
} \fi

\if0\blind
{
  \bigskip
  \bigskip
  \bigskip
  \begin{center}
    {\LARGE\bf Value of Information: sensitivity analysis and research design in Bayesian evidence synthesis}
\end{center}
  \medskip
} \fi

\bigskip

\begin{abstract} %  200 or fewer words.  194
  Suppose we have a Bayesian model which combines evidence from several different sources.  We want to know which model parameters most affect the estimate or decision from the model, or which of the parameter uncertainties drive the decision uncertainty.   Furthermore we want to prioritise what further data should be collected.    These questions can be addressed by Value of Information (VoI) analysis, in which we estimate expected reductions in loss from learning specific parameters or collecting data of a given design.  We describe the theory and practice of VoI for Bayesian evidence synthesis, using and extending ideas from health economics, computer modelling and Bayesian design.    The methods are general to a range of decision problems including point estimation and choices between discrete actions.  We apply them to a model for estimating prevalence of HIV infection, combining indirect information from several surveys, registers and expert beliefs.  This analysis shows which parameters contribute most of the uncertainty about each prevalence estimate, and provides the expected improvements in precision from collecting specific amounts of additional data. 
\end{abstract}

\noindent%
{\it Keywords:}  decision theory, research prioritisation, uncertainty %% 3-6 keywords
\vfill

\newpage
%\spacingset{1.45} % DON'T change the spacing!
\section{Introduction}
\label{sec:intro}

Bayesian modelling is a natural paradigm for decision making, in the presence of uncertainty, based on multiple sources of evidence.  However, as more data sources, parameters and assumptions are built into a model, it becomes harder to see the influence of each input or assumption.  The modelling process should involve an investigation of where the weak parts of the model are, to identify which uncertainties in the model inputs contribute most to the uncertainty in the final result or decision (\emph{sensitivity analysis}).  We might then want to assess and compare the potential value of obtaining datasets of specific designs or sizes to strengthen different parts of the model.  Furthermore, we may want to formally trade off the costs of sampling with the resulting expected improvement to decision making.   

Annual estimation of HIV prevalence in the United Kingdom has, for several years, been based on a Bayesian synthesis of evidence from various surveillance systems and other surveys \citep{goubar2008estimates,Presanis2010,DeAngelis2014a,PHEreport2016}.  This is an example of a class of problems called \emph{multiparameter evidence synthesis} \citep[e.g.][]{ades:sutton}, where the quantities of interest are not directly observable, but can be inferred from multiple indirect data sources linked through a network of model assumptions that can be expressed as a directed acyclic graph.   Markov Chain Monte Carlo is typically required to estimate the posterior. The model is typically used to inform health policies, and in this context it is important to be able to assess sensitivity to uncertain model inputs and to indicate how the model could be strengthened with further data. 

These dual aims can be achieved with \emph{value of information} (VoI) analysis, a decision-theoretic framework based on expected reductions in loss from future information.  The concepts of VoI were first set out in detail by \citet{raiffa:schlaifer}, while \citet{parmigiani2009decision} give a more recent overview.   The expected value of \emph{partial perfect information} (EVPPI) is the expected reduction in loss if the exact value of a particular parameter or parameters $\bm\theta_0$ were learnt, also interpreted as the amount of decision uncertainty that is due to $\bm\theta_0$.  The expected value of \emph{sample information} (EVSI) is the expected reduction in loss from a study of a specific design.  These concepts have been applied in various forms in three distinct areas: health economics, computer modelling and Bayesian design.

%% We present a unifying framework for VoI and extend in multiparameter evidence synthesis. 
% {\color{blue} We develop VoI tools for general multiparameter evidence synthesis, building on and bringing together three strands of literature: in health economics, computer modelling and Bayesian design.}

In health economic modelling, there is a large literature on calculation and application of VoI, see, e.g. \citet{felli1998sensitivity,willan2005value,claxton2006using,welton2008research}.  The model output in this case is the expected net benefit of each alternative policy, a known deterministic function $g(\bm\theta)$ of uncertain inputs $\bm\theta$, and the decision problem is the choice of policy that minimises $g(\bm\theta)$.  In computer modelling, see, e.g. \citet{oakley:ohagan:psa} and \citet{saltelli2004sensitivity}, the influence of a particular element $\bm\theta_0$ of $\bm\theta$ is calculated as the expected reduction in $\var(g(\bm\theta))$, if we were to learn $\bm\theta_0$ exactly.  This is equivalent to the EVPPI for $\bm\theta_0$ under a decision problem defined as point estimation of $g(\bm\theta)$ with quadratic loss \citep{oakley:ohagan:psa}.   The decision-theoretic view of Bayesian experimental design also has a long history, see, e.g. \citet{lindley1956measure,bernardo:smith,chaloner1995bayesian,berger2013statistical}, and a recent review of the computational challenges by \citet{ryan2015review}. 

However, the current tools in any one of these three areas cannot be applied directly to multiparameter evidence synthesis.  For example, it is not always feasible or desirable to make a discrete decision with a quantifiable loss, as in health economic modelling, as often the aim of an evidence synthesis is simply to estimate one or more quantities.  For a scalar quantity of interest, we might then define the ``loss'' as the posterior variance of this quantity, as in \citet{oakley:ohagan:psa}.  In computer modelling, however, tools to estimate the expected value of a proposed study to learn a particular $\bm\theta_0$ more precisely have not been developed, and it is not clear what an appropriate loss for a vector of model outputs would be.  Challenges also arise with computation.  Current methods for computing the expected variance reduction in the computer modelling field \citep{sobol2001global,saltelli2004sensitivity} assume the output is a known function $g(\bm\theta)$ of the inputs, therefore do not apply in multiparameter evidence synthesis, where MCMC is required to obtain the output.  For Bayesian design, \citet{ryan2015review} reviewed methods where evaluating the expected utility of a design (equivalent to the EVSI) is relatively inexpensive, so that maximising the utility over a complex design space is feasible.  However, this can again be difficult with MCMC.  Given a sample from the posterior $p(\theta|\x)$, potential future datasets $\y$ under a specific design can be simulated cheaply from the posterior predictive distribution, but then to obtain the expected utility, it is required to repeatedly update the posterior $p(\theta|\x,\y)$ for different $\y$, which is only feasible with Monte Carlo for smaller problems \citep[e.g.][]{han2004bayesian}.

We describe a VoI framework for sensitivity analysis and research design in evidence syntheses based on graphical models, using and extending methods from health economics, computer modelling and Bayesian design.  This is a broader class of models than those typically used in health economics or computer modelling, since the model ``output'' is not necessarily a known function of the inputs, but depends on the model parameters $\bm\theta$ and observed data $\x$ through a network of statistical models or deterministic functions, potentially with hierarchical relationships.   We apply VoI methods to the part of the HIV prevalence estimation model that estimates prevalence in men who have sex with men (MSM), in London.  Here the decision problem is point estimation of a single scalar or a vector of parameters.  We use ideas from Bayesian design to choose appropriate loss functions in this context.  We also show how methods of computing EVPPI \citep{strong:oakley:multievppi} and EVSI \citep{strong:oakley:evsi} for finite choices in health economics, based on fitting a non-parametric regression to a sample from the posterior, can be generalised to a broader class of decision problems, including point estimation.  The method for computing EVSI enables the expected utility over all potential $\y$ to be estimated cheaply without an additional level of simulation, assuming only that the information provided by $\y$ can be represented as a low-dimensional sufficient statistic $T(\y)$.

 % and not assuming the expected utility of the optimal decision given $\y$ is available analytically.
%% \citet{strong:oakley:multievppi} developed an efficient method for computing EVPPI \emph{[define earlier..]} with nonparametric regression, using only a Monte Carlo sample of $\bm\theta$ and $g(\bm\theta)$, while alternative methods exploiting the analytic structure of $g()$ are available for special cases \citep[e.g.][]{ades2004expected,madan2014strategies}.

In Section~\ref{sec:meth} we describe the general multiparameter evidence synthesis model, and define the expected value of information under different decision problems and loss functions, and in Section~\ref{sec:comp} we present methods to compute them.  In Section~\ref{sec:hiv} we describe the model for HIV prevalence estimation, and Section~\ref{sec:results} we use VoI to identify the areas of greatest uncertainty in this model and show where collecting specific data would improve the precision of the estimates of various subgroup-specific prevalences.   Finally we discuss potential extensions to the methods and application and the associated challenges.

% p139: simulate m values from post pred, calculate utility, take mean
% what's the difference? can we calc the utility for specific data?
% would need posterior to be updated given sim data, then take max NB, or post var
% instead we estimate the utilities as fitted values from regression

% is strong et al an example of one of these methods?  augmented probability model?
% augmented thing considers utility as a "likelihood" for samples, with the posterior as the "prior". sample from resulting "posterior", marginal mode of d is the optimal design
% we're fitting a second model to posterior samples.  is that "incoherent"?

\section{Theory and methods}
\label{sec:meth}

\subsection{Bayesian graphical modelling for evidence synthesis}

In our motivating applications, the general model can be represented as a directed acyclic graph (Figure~\ref{fig:dag}) in the standard way, see, e.g. \citet{lauritzen1996graphical}.  Nodes in the graph may represent scalar or vector quantities.   A set of datasets $\x = \{x_1,\ldots,x_n\}$ is observed, most generally from $n$ different sources.  These data are assumed to arise from statistical models with parameters $\mu_1,\ldots,\mu_n$ respectively, collectively denoted $\bm\mu$.  The ``founder nodes'' of the graph are denoted $\bm\phi = (\phi_1,\ldots,\phi_p)$ and given a joint prior distribution $\bm\phi \sim p(.)$ which may also include substantive information.  The full set of unknowns is denoted $\bm\theta$.  Most simply, the $\bm\mu$ could equal the $\bm\phi$ or be related to the $\bm\phi$ through deterministic functions, so that $\bm\theta=\bm\phi$.  More generally, some of the relationships in the graph could be stochastic, defining a hierarchical model, where the $\bm\mu$ themselves arise from a distribution with parameters given by the $\bm\phi$ or descendants of $\bm\phi$.  $\bm\theta$ would then comprise $\bm\phi$ and the stochastic descendants of $\bm\phi$ such as random effects. 

We further denote $\bm\alpha$ as an intermediate node in the graph, the model ``output'', which is used for decision-making.  This could be any unknown quantity, including one of the $\bm\mu$ or $\bm\phi$, a function of these, or a prediction of new data.  We may also be in a position to collect additional data, either from the same source as one of the existing datasets (e.g. $y_1$ in Figure~\ref{fig:dag}), or from a new source informing a parameter $\mu_{n+1}$ on which no direct data ($y_2$) were available.

This DAG (Figure~\ref{fig:dag}) is a generalisation of the typical structure (Figure~\ref{fig:dag:comp}) used in computer modelling \citep{oakley:ohagan:psa} where the output $\bm\alpha$ is a known (usually complicated) deterministic function of uncertain model inputs $\bm\phi$, which are given substantive priors that may be derived separately from data. 

\definecolor{deepskyblue}{rgb}{0, 0.75, 1}
\tikzstyle{node} =[minimum size = 1cm, text width=1cm, font=\normalsize, align=center]
\tikzstyle{data}    = [node, rectangle, fill=deepskyblue]
\tikzstyle{founder} = [node, circle, draw, double, line width=1.5pt, fill=yellow]
\tikzstyle{func}   = [node, circle, draw, ultra thin, fill=lightgray!5]  % same colour as background
\tikzstyle{interest}= [func, draw, line width=3pt]
\tikzstyle{datagroup}= [draw, rectangle, rounded corners, inner sep=10pt, fill=lightgray!5]
\tikzstyle{future}=[node, draw, dotted]

\begin{figure}
  \centering
  \begin{tikzpicture}[minimum width = 0.8cm, minimum height = 0.5cm]

    \node[data]               (x1)   {\(x_1\)};
    \node[right=of x1]                 (xdot) {\(\ldots\)};
    \node[data,right=of xdot] (xn)   {\(x_n\)};

    \node[func,above=of x1]  (mu1)   {\(\mu_1\)};
    \node[right=of mu1]              (mudot) {};
    \node[func,above=of xn]  (mun)   {\(\mu_n\)};

    \node[interest,above=of mudot]  (alpha)   {\(\alpha\)};

    \node[above=of alpha]                 (phidot2) {};

    \node[left=of phidot2]                 (phidot1) {\(\ldots\)};
    \node[founder,left=of phidot1]     (phi1)   {\(\phi_1\)};

    \coordinate (phi1out1) at ($(phi1) + (-60 : 2)$);
    \coordinate (phi1out2) at ($(phi1) + (-30 : 2)$);
    \node (phi1outdots) at ($(phi1) + (-45 : 2.5)$) {\(\ldots\)};
    \draw[->] (phi1) -- (phi1out1);
    \draw[->] (phi1) -- (phi1out2);

    \node[right=of phidot2]                 (phidot3) {\(\ldots\)};
    \node[founder,right=of phidot3] (phip)   {\(\phi_p\)};

    \coordinate (phipout1) at ($(phip) + (210 : 2)$);
    \coordinate (phipout2) at ($(phip) + (240 : 2)$);
    \node (phipoutdots) at ($(phip) + (225 : 2.5)$) {\(\ldots\)};
    \draw[->] (phip) -- (phipout1);
    \draw[->] (phip) -- (phipout2);

    \coordinate (ain1) at ($(alpha) + (30 : 1.3)$);
    \coordinate (ain2) at ($(alpha) + (150 : 1.3)$);
    \draw[->] (ain1) -- (alpha);
    \draw[->] (ain2) -- (alpha);

    \coordinate (aout1) at ($(alpha) + (200 : 1.2)$);
    \coordinate (aout2) at ($(alpha) + (270 : 1.2)$);
    \coordinate (aout3) at ($(alpha) + (-20 : 1.2)$);
    \node (aoutdots) at ($(alpha) + (270 : 2.5)$) {\(\ldots\)};
    \draw[->] (alpha) -- (aout1);
    \draw[->] (alpha) -- (aout2);
    \draw[->] (alpha) -- (aout3);

    \draw[->] (mu1) -- (x1);
    \draw[->] (mun) -- (xn);

    \node[future,left=of x1] (y1) {\(y_1\)};
    \node[future,right=of xn] (y2) {\(y_2\)};
    \node[func,above=of xn]  (mun)   {\(\mu_n\)};
    \node[func,above=of y2]  (munp1)   {\(\mu_{n+1}\)};
    \draw[->] (mu1) -- (y1);
    \draw[->] (munp1) -- (y2);

    \coordinate (muin1) at ($(mu1) + (45 : 1.2)$);
    \coordinate (muinn) at ($(mun) + (135 : 1.2)$);
    \coordinate (muinnp1) at ($(munp1) + (150 : 1.2)$);
    \draw[->] (muin1) -- (mu1);
    \draw[->] (muinn) -- (mun);
    \draw[->] (muinnp1) -- (munp1);

    \node (founder_key) [right = of phip, text width=3cm] {Founder nodes\\with priors};
    \node (interest_key) [below = 0.6cm of founder_key, text width=3cm] {Model output(s)\\of interest};
    \node (data_key) [below= of interest_key, right = of y2, text width=3cm] {Observed and\\future data};

    \begin{scope}[on background layer]
    \node[datagroup, fit={(phi1) (phip) (y1) (y2) (founder_key) (data_key) }] () {};
    \end{scope}

  \end{tikzpicture}
  \caption{\label{fig:dag}Directed acyclic graph for Bayesian evidence synthesis}
\end{figure}
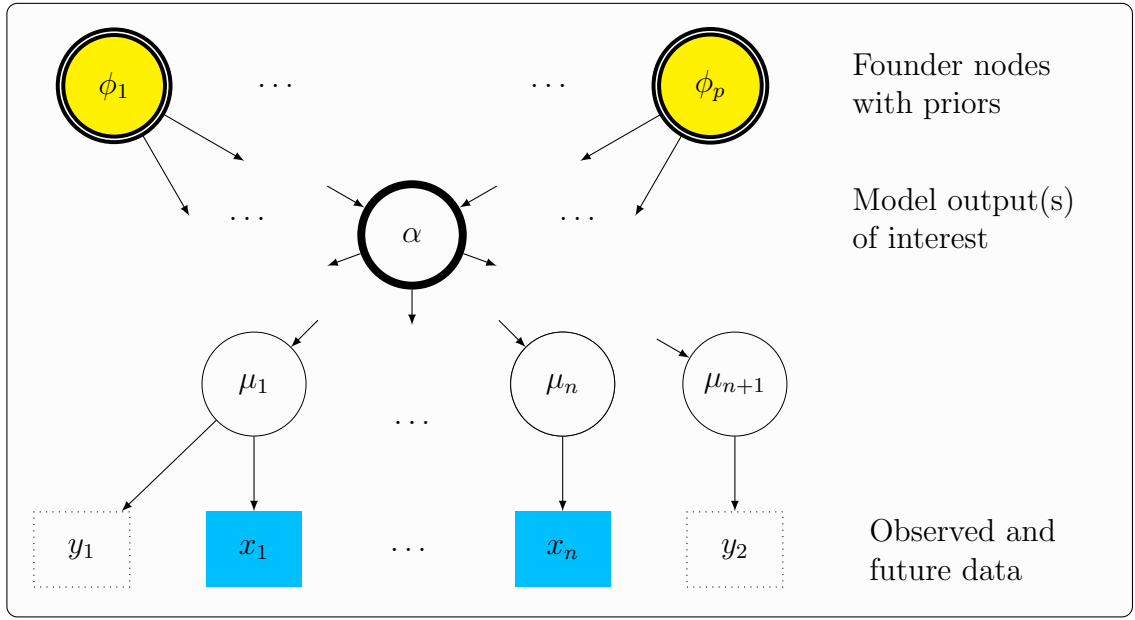

\begin{figure}
  \centering
  \begin{tikzpicture}[minimum width = 0.8cm, minimum height = 0.5cm]
    \node[founder]     (phi1)   {\(\phi_1\)};
    \node[right=of phi1]                 (phidot1) {\(\ldots\)};
    \node[right=of phidot1]                 (phidot2) {};
    \node[right=of phidot2]                 (phidot3) {\(\ldots\)};
    \node[founder,right=of phidot3] (phip)   {\(\phi_p\)};
    \node[interest,below=of phidot2]  (alpha)   {\(\alpha\)};    
    \draw[->] (phi1) -- (alpha);
    \draw[->] (phip) -- (alpha);
    \draw[->] (phidot2) -- (alpha);
    \begin{scope}[on background layer]
      \node[datagroup, fit={(phi1) (phip) (alpha) }] () {};
    \end{scope}
  \end{tikzpicture}
  \caption{\label{fig:dag:comp}Graph representing a known deterministic model}
\end{figure}
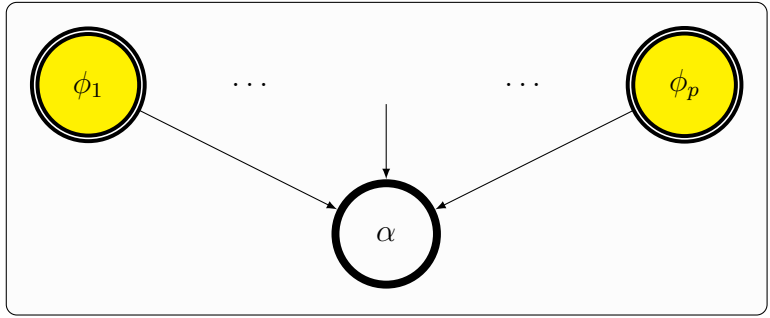

\subsection{Expected value of information: definitions}

In a general decision-theoretic framework, the purpose of the model is
to choose a decision or action $d$ from a space of possible decisions
$\mathcal{D}$, to minimise an expected loss
$E_{\bm\theta}(L(d,\bm\theta))$, with the expectation taken with
respect to the posterior distribution of $\bm\theta$.  Let
$\bm\alpha = \bm\alpha(\bm\theta)$ be the minimal subset or function
of $\bm\theta$ necessary to make the decision, so that $E_{\bm\theta}(L(d,\bm\theta)) = E_{\bm\alpha}(L(d,\bm\theta))$, $\forall d \in \mathcal{D}$.  For example, the purpose could be the choice of decision $d$ among a
finite set $\mathcal{D}=\{1,\ldots,D\}$ expected to minimise a loss
defined as a function of the parameters, so that $\bm\alpha$ would be
a vector with $D$ components $\alpha_d = f_d(\bm\theta)$ say, with
$L(d,\bm\theta) = \alpha_d$. This is the typical situation in health
policy decisions \citep[e.g.][]{claxton2006using}, where a treatment $d$
is chosen to maximise a measure of utility such as expected
quality-adjusted survival.  Alternatively, as in our examples, 
the decision could simply
be the choice of a point estimate $\hat{\bm\alpha}$ of some parameter
$\bm\alpha$, in which case the decision space $\mathcal{D}$ is the
(typically continuous) support of $\bm\alpha$ (see~\S\ref{sec:voi:specific}).

%\underset{x}{\operatorname{argmax}}

For general decision problems, let $d^* = \argmin_d E_{\bm\theta}(L(d,\bm\theta))$ be the optimal decision under \emph{current} knowledge about
$\bm\theta$, represented by the posterior distribution $p(\bm\theta|\x)$.   
Suppose now we are in a position to collect new information.  
Let $d^*_{\y}$ be the optimal decision given further
knowledge of a quantity $\y$ (either parameters or potential data) that 
informs $\bm\alpha$, so that the
updated posterior would be $p(\bm\theta|\x,\y)$. 
\begin{enumerate}
\item The \emph{expected value of perfect information} (EVPI) is
  the expected loss of the decision $d^*$ under current
  information, minus the expected loss for the decision $d^*_{\bm\alpha}$
  we would make if we knew the true $\bm\alpha$ \citep{raiffa:schlaifer}.
  \[
    E_{\bm\theta} (L(d^*,\bm\theta)) - E_{\bm\theta} (L(d^*_{\bm\alpha},\bm\theta))
  \]
  Since additional information is always expected to reduce the
  expected loss of the optimal decision
  \citep[][]{parmigiani2009decision}, the EVPI is an upper bound on
  the expected gains from any new information.

\item The \emph{expected value of partial perfect information} (EVPPI)
  for a particular (scalar or vector) parameter $\bm\phi$ is the expected reduction in loss
  if $\bm\phi$ were to be known precisely:
  \begin{equation}
    \label{eq:evppi}
    EVPPI(\bm\phi) = E_{\bm\theta} (L(d^*,\bm\theta)) - E_{\bm\phi} [  E_{\bm\theta|\bm\phi} (L(d^*_{\bm\phi},\bm\theta)) ]
  \end{equation}
  where $d^*_{\bm\phi}$ is the optimal decision if $\bm\phi$ were known.  This is an upper bound on the potential value of data $\y$
  which inform only $\bm\phi$.  In a graphical model, this means data
  $\y$ that are conditionally independent of $\bm\theta$
  given $\bm\phi$, for example $\y=y_1$ and $\bm\phi = \mu_1$ in
  Figure~\ref{fig:dag}.
  
%%% markov blanket:  A is cond indep of all nodes outside Markov blanket, given its Markov blanket.

\item The \emph{expected value of
  sample information} $EVSI(\y)$ is the reduction in loss we would
expect from collecting an additional dataset $\y$ of a specific design.
\begin{equation}
  \label{eq:evsi}
  EVSI(\y)  =  E_{\bm\theta}(L(d^*, \bm\theta)) - E_{\y} \left[  E_{\bm\theta|\y} (L(d^*_{\y}, \bm\theta) \right]
\end{equation}
The inner expectation is now with respect to the updated posterior
distribution of $\bm\theta|\y$, after learning $\y$ as well as the
existing data $\x$, or ``preposterior'' \citep{berger2013statistical}.
If we can express the costs $C(\y)$ of obtaining $\y$ using the same
loss metric, we can further define the \emph{expected
  net benefit of sampling} as $EVSI(\y) - C(\y)$, and typically seek
the sample size that maximises this \citep{parmigiani2009decision}.

% clemen and reilly 2001 call sample information ``imperfect information''

\end{enumerate}

\subsection{Value of information in different decision problems}
\label{sec:voi:specific}

\paragraph{Finite-action decisions}
 For a choice of $d$ among a finite set $\{1,\ldots,D\}$
with loss $L(d,{\bm\theta}) = \alpha_d$ and $\bm\alpha = \{\alpha_1,\ldots,\alpha_d,\ldots,\alpha_D\}$, the expected loss
  with current information is $\min_d \{ E_{\bm\alpha}(\alpha_d) \}$, so \citep{raiffa:schlaifer}
  \begin{eqnarray*}
EVPI  & = & \min_d \{ E_{\bm\alpha}(\alpha_d) \} - E_{\bm\alpha} \min_d\{\alpha_d\}\\
EVPPI(\bm\phi)  & = & \min_d \{ E_{\bm\alpha}(\alpha_d) \} - E_{\bm\phi} \min_d\{E_{\bm\theta|\bm\phi}(\alpha_d)\} \\
EVSI(\y)  & = & \min_d \{ E_{\bm\alpha}(\alpha_d) \} - E_\y \min_d\{E_{{\bm\alpha}|\y}(\alpha_d)\}
  \end{eqnarray*}

\paragraph{Point estimation of a parameter}
When the decision is the choice of a point estimate $\hat{\bm\alpha}$
  of a vector of parameters $\bm\alpha$, with quadratic loss
  \begin{equation}
    \label{eq:quadloss}
  L(\hat{\bm\alpha},\bm\alpha) = (\hat{\bm\alpha} - \bm\alpha)^T H (\hat{\bm\alpha} -
  \bm\alpha)
  \end{equation}
  for a symmetric, positive-definite $H$, the optimal estimate with current information is the
  posterior mean, $\hat{\bm\alpha} = E_{\bm\alpha}(\bm\alpha)$.  For a scalar $\bm\alpha=\alpha$ and $H=1$, the
  expected loss is $\var(\alpha)$ under current information and zero
  under perfect information, so that $EVPI=\var(\alpha)$ and
  \begin{eqnarray}
    EVPPI(\bm\phi) & =  & \var(\alpha) - E_{\bm\phi}\left[\var_{\alpha|\bm\phi}(\alpha | \bm\phi)\right] \label{eq:evppi:var}\\
    EVSI(\y) & = &  \var(\alpha) - E_{\y}\left[\var_{\alpha|\y}(\alpha | \y)\right] \label{eq:evsi:var}
  \end{eqnarray}
  the expected reduction in variance given new information.
  Expression (\ref{eq:evppi:var}) is used by \citet{oakley:ohagan:psa} and
  \citet{saltelli2004sensitivity} as a measure of sensitivity of the
  output of a deterministic model $\alpha = g(\phi,\ldots)$ to an
  uncertain input $\phi$, termed the \emph{main effect} of $\phi$, but
  this has not been extended to the EVSI of a potential
  dataset $\y$ in this context.

  Alternatively, an absolute error loss \citep{bernardo:smith} gives
  $\hat\alpha$ as the posterior median and value measures based on the
  mean absolute deviation.

\paragraph{Point estimation of multiple parameters}
The purpose of a multiparameter evidence synthesis of the form in Figure~\ref{fig:dag} is typically to estimate several correlated parameters of interest, comprising a vector $\bm\alpha$, say.  Most simply, we could conduct independent value of information analyses for each component of $\bm\alpha$.  In more formal decision analyses we may want a scalar loss for the overall vector $\bm\alpha$.  There are various alternatives based on generalisations $v(\bm\alpha)$ of the variance, which can be used instead of the scalar variance $\var(\alpha)$ in equations (\ref{eq:evppi:var})--(\ref{eq:evsi:var}) to define the expected value of information.  These have been applied in the context of Bayesian study design, and we show how they can also be used for the EVPPI and EVSI in evidence synthesis models.

\begin{enumerate}
\item If $H=\cc \cc^T$ in the quadratic loss~(\ref{eq:quadloss}), for some vector of weights $\cc$, then the expected loss is $v(\bm\alpha) = \cc^T \cov(\bm\alpha) \cc = \var(\cc^T \bm\alpha)$, corresponding to optimal estimation of the weighted sum of the parameters, $\cc^T\bm\alpha$.   For example, when the elements $\alpha_s$ of $\bm\alpha$ are weighted equally, the goal is to minimise the sum of all elements $(r,s)$ of the covariance matrix, $v(\bm\alpha) = \sum_{r,s}\cov(\bm\alpha)_{r,s}$, or, if the $\alpha_s$ are also independent of each other, $v(\bm\alpha) = tr(\cov(\bm\alpha)) = \sum_s \var(\alpha_s)$.  The same \emph{absolute} reductions in variance for different components of $\bm\alpha$ would then be valued equally.  More generally, if $\cc$ is given a prior, then loss~(\ref{eq:quadloss}) also arises (see~\citet{chaloner1995bayesian} and references therein).  Designs that minimise~(\ref{eq:quadloss}) are Bayesian analogues of classical \emph{A-optimal} designs. See also \citet{lamboni2011multivariate} for similar measures of sensitivity for multivariate outputs in deterministic computer models.  %% and \citet{gamboa2013sensitivity} 

\item A Bayesian \emph{D-optimal} design, on the other hand, minimises the \emph{determinant} $v(\bm\alpha) = \det(\cov(\bm\alpha))$ \citep{chaloner1995bayesian,ryan2015review}.   This simplifies to the \emph{product} of the $\var(\alpha_s)$ when the $\alpha_s$ are independent and equally-weighted.   Equivalently, a standardised version $\det(\cov(\bm\alpha))^{1/S}$, where $S$ is the number of components of $\bm\alpha$, represents a geometric average variance of the $\alpha_s$, adjusted for their covariance. 
  
Here the same \emph{relative} reductions in variance for different components of $\bm\alpha$ would then be valued equally, which would be more appropriate when the output of interest $\bm\alpha$ comprises quantities on very different scales and/or with different interpretations.

\end{enumerate}

\section{Computation of value of information}
\label{sec:comp}

\subsection{Partial perfect information}
\label{sec:evppi:comp}

Computation of the EVPPI in general is not straightforward.  Given a sample from the posterior distribution, the first term in~(\ref{eq:evppi}) can be calculated by a Monte Carlo mean.  The double expectation in the second term is more challenging. \citet{strong:oakley:multievppi} presented a method for estimating the EVPPI which avoids an expensive nested Monte Carlo procedure.  However this only applied to finite-choice decision problems.  We extend the scope of this method to more general problems, including point estimation. Suppose that, given a state of knowledge about $\bm\alpha$ represented by a distribution $\psi(.)$, the expected loss under the optimal decision is a known function $h$ of the mean of $\bm\alpha$ under that distribution.
\begin{equation}
  \label{eq:knownfun}
  E_{\psi} (L(d^*_\psi,\bm\theta)) = h(E_{\psi}(\bm\alpha)).
\end{equation}
If $\psi(.)$ is the current posterior, this is $h(E_{\bm\alpha}(\bm\alpha))$, and if we were to learn the value of $\bm\phi$, the expected loss would be $h(E_{\bm\alpha|\bm\phi}(\bm\alpha|\bm\phi))$.  We can estimate $E_{\bm\alpha|\bm\phi}(\bm\alpha|\bm\phi)$ by expressing
\begin{equation}
  \label{eq:evppi:reg}
  \bm\alpha  =  E_{\bm\alpha|\bm\phi}(\bm\alpha|\bm\phi) + \epsilon  =  g(\bm\phi) + \epsilon
\end{equation}
where $\epsilon$ is an error term with mean zero.    Then using a Monte Carlo sample of $(\bm\alpha^{(k)}, \bm\phi^{(k)}): k=1,\ldots,K$, we estimate $g(\bm\phi)$ by regression of $\bm\alpha$ on $\bm\phi$.  If $\bm\phi$ comprises $p$ parameters that could be learnt simultaneously, the regression will have $p$ predictors.  Since the functional form of $g()$ will not be known in general, nonparametric regression methods are preferred.   This produces a fitted value $\hat g(\bm\phi^{(k)})$ for each $k$.

Then the second term in~(\ref{eq:evppi}) is estimated by a Monte Carlo mean
\[
E_{\bm\phi} [  E_{\bm\theta|\bm\phi} (L(d^*_{\bm\phi},\bm\theta)) ] = E_{\bm\phi} [  h (E_{\bm\alpha|\bm\phi}(\bm\alpha|\bm\phi)) ] \approx \frac{1}{K}\sum_{k=1}^K h(\hat g(\bm\phi^{(k)})).
\]
\citet{strong:oakley:multievppi} only presented this method for finite choices, where $\bm\alpha$ is a vector and $h(E(\bm\alpha)) = \max_d \{E(\alpha_d)\}$.   Then a separate $g_d()$ is estimated to relate each $\alpha_d$ to $\bm\phi$, and
\[E_{\bm\phi} [  E_{\bm\theta|\bm\phi} (L(d^*_{\bm\phi},\bm\theta)) ] \approx \frac{1}{K}\sum_{k=1}^K \max_d \{\hat g_d(\bm\phi^{(k)})\},
\]
Our more general formulation of this algorithm, which expresses the optimal loss as $h(E(\bm\alpha))$, can be used for point estimation problems.  For estimation of a scalar $\alpha$ with quadratic loss, $h(E_{\alpha}(\alpha)) = E[(\alpha - E_\alpha(\alpha))^2] = \var(\alpha)$.  We estimate $\var(\alpha|\bm\phi^{(k)})$ by the squared residual $(\alpha - \hat g(\bm\phi^{(k)}))^2$, substitute this for $h(\hat g(\bm\phi^{(k)}))$ and estimate $E_{\bm\phi}\left[\var_{\alpha|\bm\phi}(\alpha | \bm\phi)\right]$ as the mean, over $k$, of the squared residuals.  Equivalently we can estimate $\var(\theta) - E_{\bm\phi}\left[\var_{\alpha|\bm\phi}(\alpha|\bm\phi)\right] = \var_{\bm\phi}(E_{\alpha|\bm\phi}(\alpha|\bm\phi))$ as the variance, over $k$, of the fitted values.  Similarly, for vector $\bm\alpha$ and loss functions based on $\cov(\bm\alpha)$, we can fit regressions to get the marginal mean for each component $\alpha_d$, and calculate the empirical covariance matrix of the residuals.

Several methods of nonparametric regression have been suggested.  For small $p$, \citet{strong:oakley:multievppi} used generalized additive models, with tensor products of spline smoothers to represent interactions between different components of $\bm\phi$.  Where $\bm\phi$ included about $p=5$ or more components, Gaussian process regression was recommended as a more efficient way of modelling interactions, though the resulting matrix computations rapidly become impractical as the MCMC sample size $K$ increases.  \citet{heath2015efficient} developed an integrated nested Laplace approximation for fitting Gaussian processes more efficiently in this context where $p>=2$.  For the application in Section~\ref{sec:hiv} (with $K=150000$, $p\leq 3$), we have found multivariate adaptive regression splines~\citep{friedman1991multivariate} via the \emph{earth} R package~\citep{earthpackage} to be more efficient.    Standard errors for the EVPPI estimates can be calculated in general by simulating from the asymptotic normal distribution of the regression coefficients \citep{mandel:simci}.

For expected losses which are functions of the median or other quantiles, such as absolute error loss, a similar method based on nonparametric quantile regression could be devised.

\subsection{Sample information}
\label{sec:evsi:comp}

The regression method above can also be used to estimate the expected value of sample information $EVSI(\y)$.  \citet{strong:oakley:evsi} described the method for finite decision problems.  Again we generalize this to any problem satisfying condition~(\ref{eq:knownfun}), including point estimation.  The method requires that the information provided by the data $\y$ can be expressed as a low-dimensional sufficient statistic $T(\y)$, so that $E_{\alpha|\y}(\bm\alpha | \y) = E_{\alpha|\y}(\bm\alpha | T(\y))$.  This could be a point estimator of the parameter $\mu$ (as in Figure~\ref{fig:dag}) that $\y$ gives direct information on.  As in~(\ref{eq:evppi:reg}), we can write
\[
\bm\alpha  =  E_{\bm\alpha|\y}(\bm\alpha|T(\y)) + \epsilon  =  g(T(\y)) + \epsilon
\]
and estimate $g()$ using a regression fitted to a Monte Carlo sample of $(\bm\alpha^{(k)}, T(\y^{(k)})): k=1,\ldots,K$, where $\y^{(k)}$ are drawn from their posterior predictive distribution.  Then the fitted values $\hat g(T(\y^{(k)}))$ enable the double expectation to be estimated as
\[
E_{\y} [  E_{\theta|\y} (L(d^*_{\y},\bm\theta)) ] = E_{\y} [  h (E_{\alpha|\y}(\bm\alpha|\y)) ] \approx \frac{1}{K}\sum_{k=1}^K h(\hat g(T(\y^{(k)}))).
\]
Then, for example, for point estimation with quadratic loss, this is the estimated residual variance from the regression, as in Section~\ref{sec:evppi:comp}.

\section{HIV prevalence estimation model}
\label{sec:hiv}

We consider the sub-model of the full HIV burden model \citep{DeAngelis2014a,PHEreport2016} that estimates HIV prevalence in men who have sex with men (MSM), in London.  We examine two subgroups of MSM: those who have attended a genitourinary medicine (GUM) clinic in the past year (GMSM) and those who have not (NGMSM), denoting the proportion of all men who are in these subgroups by $\rho_G$ and $\rho_N$ respectively.   For each group $g \in (G,N)$, we aim to estimate simultaneously these subgroup proportions $\rho_g$, prevalence of HIV in this group $\pi_g$ and the proportion of infections that are diagnosed, $\delta_g$. Given these parameters, further important quantities are easily derived: the prevalence of diagnosed ($\pi_g\delta_g = \pidelta_g$) and undiagnosed ($\pi_g(1-\delta_g) = \pinodelta_g$) infection; and the numbers of MSM living with diagnosed ($\mu_{Dg} = \mu_{pop}\rho_g\pidelta_g$) and undiagnosed ($\mu_{Ug} = \mu_{pop}\rho_g\pinodelta_g$) infection, where $\mu_{pop}$ is the number of men (MSM and non-MSM) living in London.  Parts of the model refer to a third subgroup, previous MSM (PMSM), men who no longer have sex with men, but the prevalence among this group is much lower, and we do not describe this part in detail.

\begin{figure}[p]
  \centering
  \scalebox{1.3}{\rotatebox{90}{
    \input{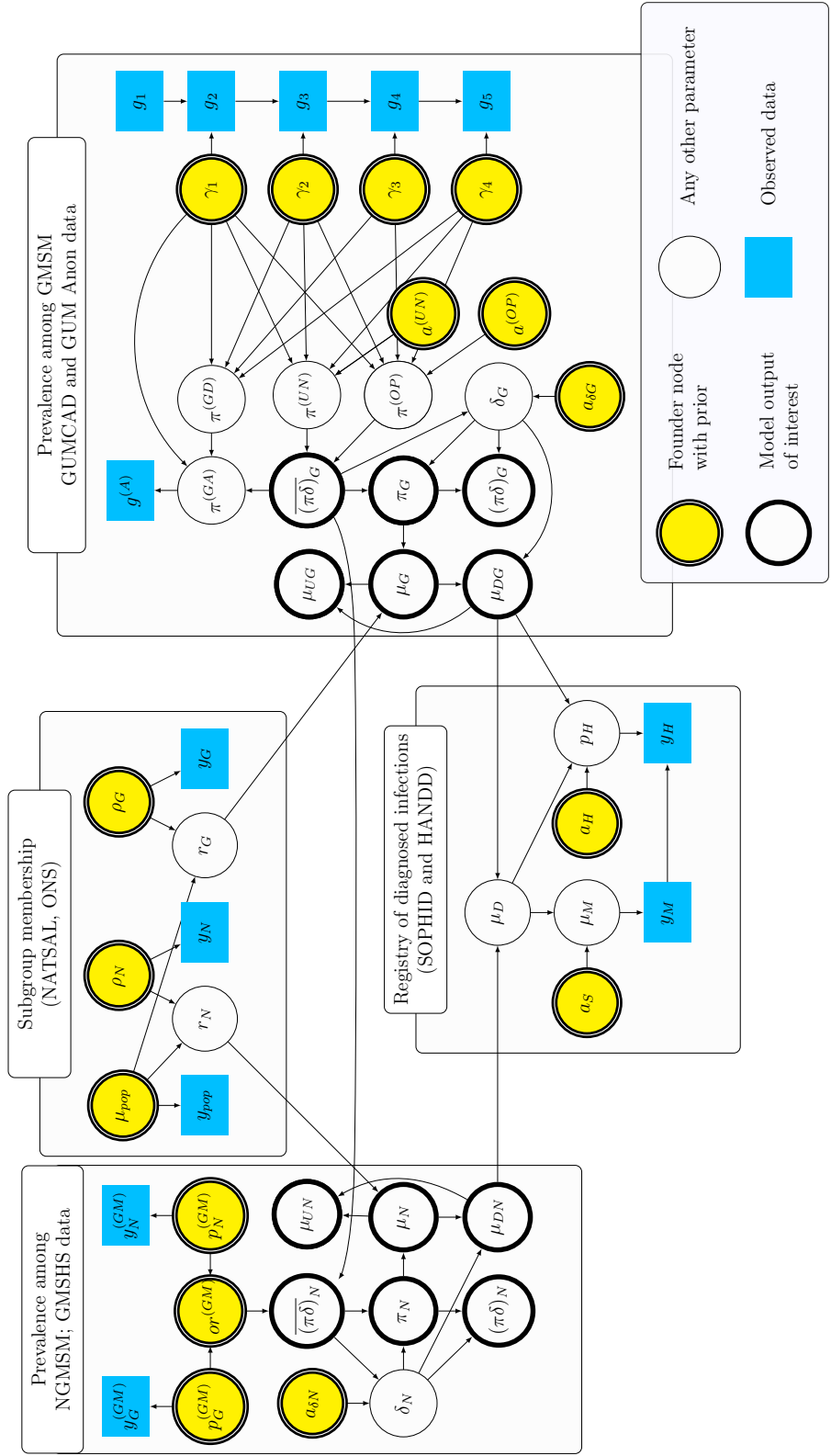}
  }}
\caption{\label{fig:hiv:dag}Directed acyclic graph for HIV prevalence estimation model.}
\end{figure}

We construct a Bayesian model to link these quantities with the available evidence provided by various routinely-collected and survey datasets as well as expert belief.  Figure~\ref{fig:hiv:dag} shows a directed acyclic graph representing this model, in the form of Figure~\ref{fig:dag}, distinguishing founder nodes, observed data, and outputs of interest.  The following sections explain in detail the quantities and relationships illustrated in Figure~\ref{fig:hiv:dag}.   All data and estimates refer to the year 2012 (unless indicated) and the Greater London area.  

%% Dotted lines indicate dependencies included or excluded in alternative model scenarios.  ????

%% FOR DIAGNOSED, need link to pideltag from gamma1 and pigd = product of gammas = prev of newly diagnosed inf
%% prod of gammas is p(no prev diag)*p(offer)*p(accept)*p(pos), doesn't use strong assumptions of prev among unoffered/refusers, but does it assume pops the same

%% FOR CUTTING UNDIAGNOSED - remove arrows from piun, piop to pinodeltag

\subsection{Subgroup membership}
\label{sec:natsal}
The total male population of London, $\mu_{pop}$, is informed by published data $y_{pop}$ \citep{ONS2012}, assumed to be a Poisson count: $y_{pop} \sim Po(\mu_{pop})$. The estimated number of people in each group $g$ is therefore $r_g = \rho_g \mu_{pop}$, where we assume a prior $\log(\mu_{pop}) \sim N(0,1000^2)$. Estimates of the subgroup proportions $\rho_g$ are informed by data from the National Survey of Sexual Attitudes and Lifestyles \citep{NATSAL3}: \iftoggle{nodata}{$y_G, y_N, y_P$}{$y_G=7$, $y_N=38,y_P=10$}, out of $y_{NAT}=824$ men, which we assume to come from a multinomial distribution with probabilities $\rho_G,\rho_N,\rho_P$ given a uniform Dirichlet prior. $P$ here refers to the PMSM subgroup, with $\rho_P$ being the corresponding proportion of men in the group.     Thus the expected number of people with HIV (diagnosed or undiagnosed) in group $g$ is $\mu_g = \pi_g r_g$.

\subsection{Registry of diagnosed infections and diagnosed prevalence}
%% SOPHID updated quarterly
Individuals diagnosed with HIV and accessing care in the UK are
reported to the SOPHID registry (Surveillance of Prevalent HIV
Infections Diagnosed) \citep{PHEreport2016}.  From SOPHID we
obtain the reported number of HIV diagnoses for MSM, $y_M \sim
Po(\mu_M)$.  We assume a small reporting bias of unknown direction,
through $\log(\mu_M) = a_S + log(\mu_D)$ where $\exp(a_S) \sim
N(1,0.018^2)$, giving a prior $90\%$ interval of about $(-3\%,3\%)$
for the adjustment to the number of MSM HIV diagnoses $\mu_M$.  After
this adjustment, $\mu_{D} = \mu_{DG} + \mu_{DN} + \mu_{DP}$ is the
expected number of diagnoses among MSM, summed from the expected
numbers of diagnoses among GMSM, NGMSM and PMSM respectively.
The following sections explain where $\mu_{DG},\mu_{DN}$
come from; $\mu_{DP}$ is modelled using similar techniques. 

Since SOPHID does not record GUM clinic attendance, to strengthen the evidence on diagnosed prevalence in GMSM we include data from the HIV and AIDS New Diagnoses Database (HANDD) \citep{PHEreport2016}, recording how many of the $y_{M}$ prevalent diagnosed MSM were newly diagnosed in 2012 and reported to have been diagnosed initially in a GUM clinic. These new diagnoses, denoted $y_{H}$, are modelled as $y_{H} \sim Bin(y_{M}, p_{H})$, where $p_H$ is assumed to be a lower bound for the proportion of prevalent diagnosed MSM who have attended a GUM clinic in 2012. This bound is expressed through $p_{H} = a_H \mu_{DG} / \mu_{D}$, where $a_H \sim U(0,1)$ is the unknown probability that a prevalent diagnosed MSM who has attended a GUM clinic in 2012 was newly diagnosed that year. $y_H$ therefore gives us additional indirect information on $\mu_{DG}$, the number of prevalent diagnosed GMSM.

The number of \emph{diagnosed} infections is related to the total
number of infections in each group $g=G,N$ as $\mu_{Dg} = \delta_g
\mu_{g}$.  The proportion of infections that are diagnosed $\delta_g$
is not known, but given our inferences about the undiagnosed
prevalence $\pinodelta_g = \pi_g(1 - \delta_g)$ (explained in the
subsequent sections), we can exploit the implicit constraint $1 -
\delta_g > \pinodelta_g$.  Therefore we define $\delta_g = a_{\delta
  g}(1 - \pinodelta_g)$, with $a_{\delta g} \sim U(0,1)$, and the
diagnosed prevalence $\pidelta_g = \pi_g\delta_g$ in each group follows.

\subsection{Undiagnosed prevalence among GMSM}
\label{sec:gumcad}
%% GUMCAD for England
Information about undiagnosed infections in GMSM is obtained from GUMCAD (Genitourinary Medicine Clinic Activity Dataset) \citep{PHEreport2016} a registry of attendance episodes in GUM clinics.  HIV tests are offered routinely to previously-undiagnosed patients.   Thus we have a sequence of observations $g_i$, representing firstly the number of GUM clinic visits ($g_1=35121$) and then the number of these where the patient has no previous HIV diagnosis ($g_2\iftoggle{nodata}{}{=34187}$), an HIV test is offered ($g_3\iftoggle{nodata}{}{=30570}$), an HIV test is accepted ($g_4\iftoggle{nodata}{}{=29529}$), or a new HIV diagnosis is made ($g_5\iftoggle{nodata}{}{=855}$).
For $i=2,\ldots,5$, $g_i \sim Bin(g_{i-1}, \gamma_{i-1})$, with priors $\gamma_1,\gamma_2,\gamma_3 \sim U(0,1)$ and $\gamma_4 \sim U(0,0.15)$ (see below).  An HIV infection may therefore remain undiagnosed if either a test is not offered or the patient opts out of testing.   We can then decompose the prevalence of undiagnosed infection $\pinodelta_G$ into ``unoffered'' $\pi^{(UN)}$ and ``opt-out'' $\pi^{(OP)}$ components.  
\begin{equation}
  \label{eq:gumcad:undiag}
\pinodelta_G = \pi^{(UN)} + \pi^{(OP)}.  
\end{equation}
Both of those require strong prior assumptions to estimate, which will later be relaxed in a sensitivity analysis (\S\ref{sec:alternative}).  Firstly, the prevalence of infection that remains undiagnosed due to an unoffered test is
\[ \pi^{(UN)} = \gamma_1(1 - \gamma_2)p^{(UN)} \]
where $\gamma_1(1 - \gamma_2)$ is the proportion of clinic attenders that
are undiagnosed but not offered a test, and
$p^{(UN)}$ is the probability that a test would be positive for these
people.  We assume the prevalence in this group is between 0.5 and 1.5
times the prevalence in people actually tested, and
$\logit(p^{(UN)}) = \logit(\gamma_4) + a^{(UN)}$, with
$a^{(UN)} \sim U(\log(0.5), \log(1.5))$

Secondly, the prevalence of infection remaining undiagnosed due to refusing a test is
\[ \pi^{(OP)} = \gamma_1\gamma_2(1 - \gamma_3)(\gamma_4 + a^{(EX)}) \]
$\gamma_1\gamma_2(1 - \gamma_3)$ is the proportion of clinic attenders
that are undiagnosed and offered a test but opt out.  We
assume this group has an underlying HIV prevalence higher than those
given tests, but not more than 15\%, so that the
excess prevalence in this group is $a^{(EX)} = a^{(OP)}(0.15 - \gamma_4)$, 
where $a^{(OP)} \sim U(0,1)$, and the prior on $\gamma_4$ is truncated above at 0.15.   

A small amount of additional evidence on $\pinodelta_G$ is available from another dataset, GUM Anon, a convenience survey of men not previously diagnosed with HIV who had attended a GUM clinic in the previous year.   This gives direct information about the prevalence of HIV among previously undiagnosed GMSM, 
\begin{equation}
  \label{eq:gumanon}
  \pi^{(GA)} = (\pinodelta_G + \pi^{(GD)}) / \gamma_1,  
\end{equation}
where $\pi^{(GD)} = \prod_1^{4}\gamma_r$ is the prevalence of newly-diagnosed infection among clinic attenders.  The data in GUM Anon are $g^{(A)} \sim Bin(g^{(AN)}, \pi^{(GA)})$, where \iftoggle{nodata}{}{$g^{(A)}=4$ and }$g^{(AN)}=85$. 
% we use 2012 version, also one in 2013. Also gives concealment (presence or absence in sample of ARV).

\subsection{Undiagnosed prevalence among NGMSM}
\label{sec:gmshs}

To inform undiagnosed HIV prevalence in NGMSM, we use data from the Gay Men's Sexual Health Survey (GMSHS) \citep{GMSHS}, based on face-to-face interviews in selected venues where participants were offered anonymous HIV tests.  While this group is likely to have a higher HIV prevalence than the general population, we assume that the \emph{relative odds} of having HIV between NGMSM and GMSM is the same as in the general population.  The GMSHS data provide the numbers $y^{(GM)}_g$ out of $n^{(GM)}_g$ previously-undiagnosed people in group $g$ who tested positive for HIV (\iftoggle{nodata}{}{20 out of }493 GMSM and \iftoggle{nodata}{}{20 out of }452 NGMSM) so that $y^{(GM)}_g \sim Bin(n^{(GM)}_g,p^{(GM)}_g)$, with $p^{(GM)}_g \sim U(0,1)$.  Defining the odds $o(p) = p/(1-p)$, we apply the resulting odds ratio $or^{(GM)} = o(p^{(GM)}_N) / o(p^{(GM)}_G)$ to the baseline estimated from GUMCAD (Section~\ref{sec:gumcad}), giving $o(\pinodelta_N)$ = $o(\pinodelta_G) or^{(GM)}$.

\subsection{Alternative assumptions} 
\label{sec:alternative} 

The results presented in section~\ref{sec:results} are for the above model assumptions, unless specified otherwise.  Two alternative assumptions are also explored.
\begin{description}
\item[(a) Undiagnosed prevalence from GUM Anon only]
To avoid the strong prior assumptions on prevalence among those not offered a test or refusing a test, which are necessary to use the GUMCAD data to infer $\pinodelta_g$, we could infer $\pinodelta_g$ from GUM Anon alone.  To construct this model, we replace equation~(\ref{eq:gumcad:undiag}) by a $U(0,1)$ prior on $\pinodelta_g$, although the GUMCAD data are still used to estimate the parameters $\pi^{(GD)}$ and $\gamma_1$ relating the prevalence in GUM Anon to $\pinodelta_g$.
\item[(b) GUMCAD also informs diagnosed prevalence]
  Instead of being inferred indirectly though the graph, the diagnosed prevalence can be modelled directly as 
  \begin{equation}
    \label{eq:gumcad:diag}
    \pidelta_G = (1 - \gamma_1) + \gamma_1\gamma_2\gamma_3\gamma_4,     
  \end{equation}
where $1-\gamma_1$ is the probability of a previous diagnosis, and $\gamma_1\gamma_2\gamma_3\gamma_4$ is the probability of newly-diagnosed infection, in GUMCAD.  This is not done in the base case due to concerns about inconsistencies in reporting of diagnoses between GUMCAD and SOPHID/HANDD. 
\end{description}

\section{Value of information in HIV prevalence model}
\label{sec:results}

The model outputs of interest (as in Figures~\ref{fig:dag},\ref{fig:hiv:dag}) are $\bm\alpha = $($\pidelta_G$, $\pidelta_N$, $\pinodelta_G$, $\pinodelta_N$, $\mu_{DG}$, $\mu_{DN}$, $\mu_{UG}$, $\mu_{UN}$, $\mu$), the diagnosed and undiagnosed prevalences among both GMSM and NGMSM, and the corresponding absolute numbers of people living with HIV/AIDS (or ``case-counts''), and the total number of MSM with HIV/AIDS $\mu=\mu_{DG}+\mu_{DN}+\mu_{UG}+\mu_{UN}$.   Samples from the posterior distributions are generated using Hamiltonian Monte Carlo methods in the Stan software \citep{stan:manual}.  These are illustrated in Figure~\ref{fig:res:prev} along with the overall prevalence $\pi_g=\pidelta_g+\pinodelta_g$ in each group $g$, and each of these quantities summed over the two groups $g$.   The estimates of diagnosed prevalence in all MSM (top panel) are reasonably precise, while the corresponding estimates for NGMSM and GMSM are more uncertain.  Estimates of undiagnosed prevalence are lower and more precise.  Full results under the two alternative assumptions are presented in the appendix.

\begin{figure}

\begin{knitrout}
\definecolor{shadecolor}{rgb}{0.969, 0.969, 0.969}\color{fgcolor}
\includegraphics[width=\maxwidth]{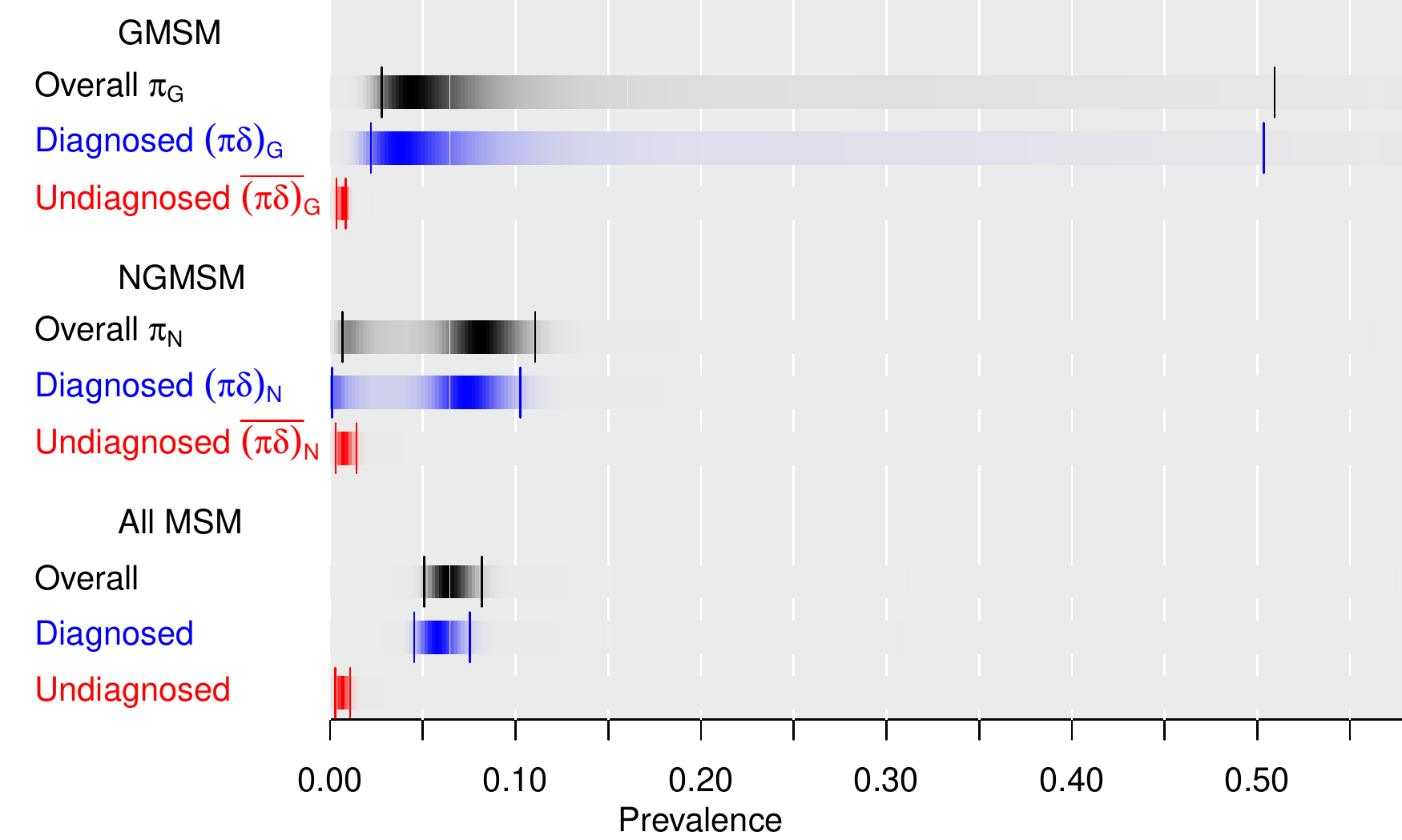} 

\includegraphics[width=\maxwidth]{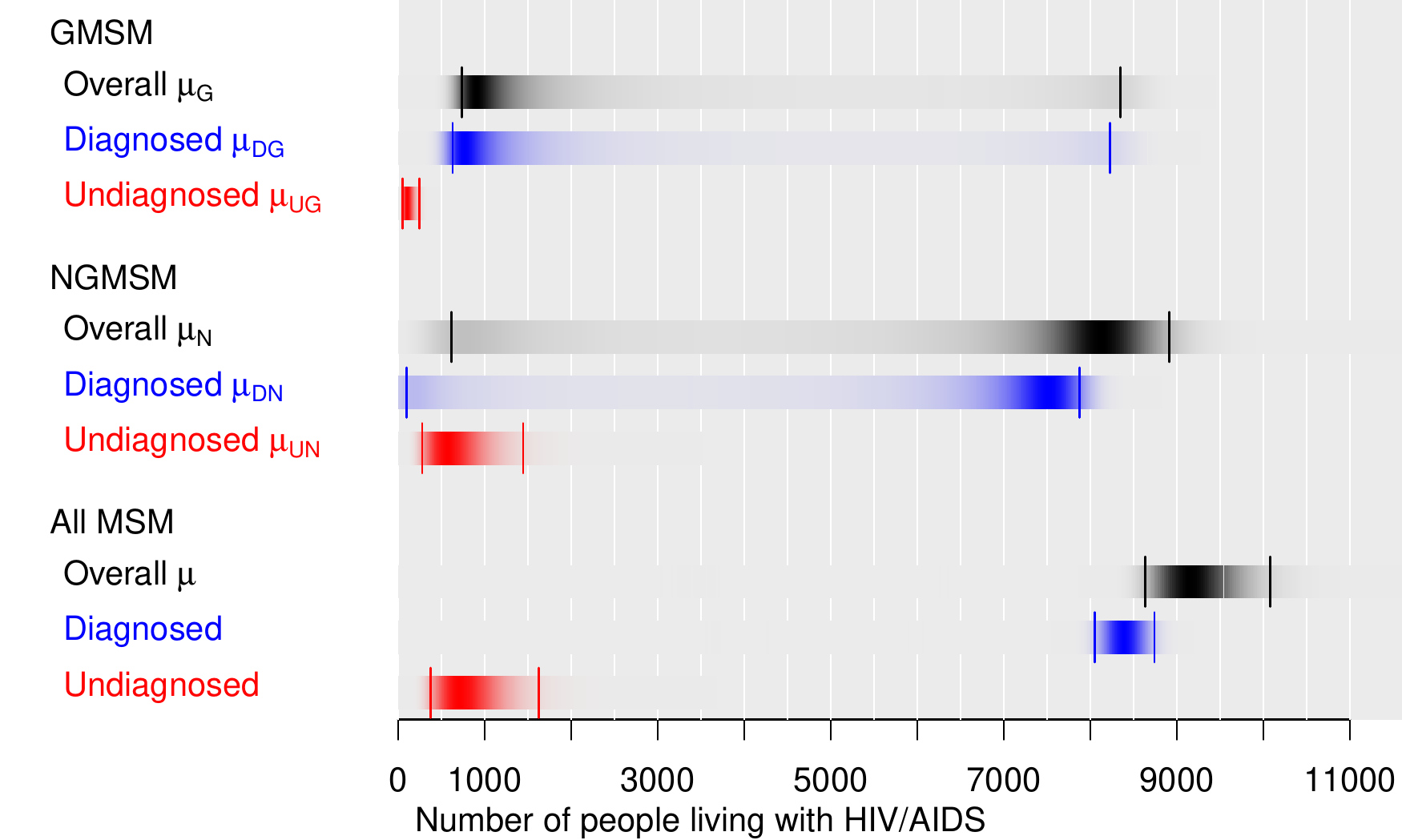} 

\end{knitrout}
  \caption{Posterior distributions of HIV prevalence (top) and numbers of MSM living with HIV/AIDS (bottom), London 2012. Darkness within each strip proportional to posterior density, with 95\% credible intervals indicated.}
  \label{fig:res:prev}
\end{figure}

\subsection{Partial perfect information (EVPPI) for single outputs}

Defining the decision problem as point estimation of $\bm\alpha$ with quadratic loss, we use EVPPI formula~(\ref{eq:evppi:var}) to determine which parameters $\phi$ contribute most to the uncertainty about each component of $\bm\alpha$, thus which $\phi$ may be worth learning more precisely.   We will take $\phi$ to include the founder nodes of the graph illustrated in Figure~\ref{fig:hiv:dag}.  Since they are related to the $\bm\alpha$ through a network of deterministic functions, perfect knowledge of these implies perfect knowledge of $\bm\alpha$.  Each of the $\phi$ are either directly informed by data or given a substantive prior distribution based on belief.   In the former case, EVPPI measures the maximum potential value of collecting more data from the same source.   In the latter case, it will not necessarily be feasible to collect data to improve the precision of the belief, but EVPPI is still useful as a measure of how much of the uncertainty in $\bm\alpha$ is explained by the uncertainty in the parameter.

The results are presented in Figure~\ref{fig:res:evppi} as a grid whose $r,s$ entry is colored according to $EVPPI_{\alpha_s}(\phi_r) / \var(\alpha_s)$, the proportion of variance in $\alpha_s$ which would be reduced if we learnt $\phi_r$.  The lighter cells correspond to $\phi_r$ with greater EVPPI.  Standard errors in these and all following EVPPI and EVSI estimates, arising from uncertainty in the coefficients of the regression~(\ref{eq:evppi:reg}), were negligible, at less than 1\% of the EVPPI or EVSI estimates.

\begin{figure}
\begin{knitrout}
\definecolor{shadecolor}{rgb}{0.969, 0.969, 0.969}\color{fgcolor}
\includegraphics[width=\maxwidth]{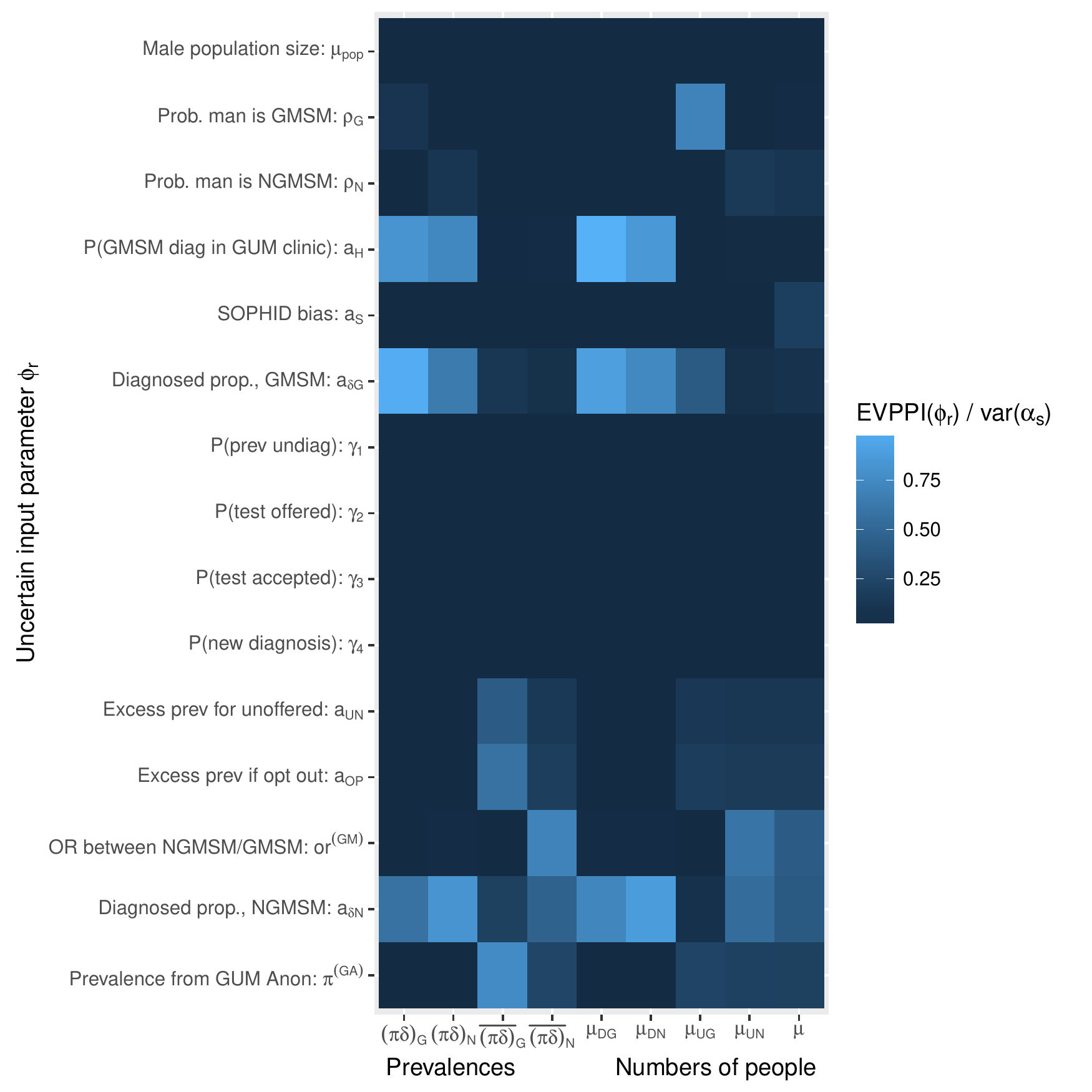} 

\end{knitrout}
  \caption{Expected value of partial perfect information in the HIV prevalence model.}
  \label{fig:res:evppi}
\end{figure}

The parameters $a_{\delta G}$ and $a_{\delta N}$, governing the proportions of HIV infections in each group that are diagnosed in each of the two groups, and the probability $a_H$ that a GMSM is newly diagnosed in a GUM clinic, explain most of the uncertainty in the \emph{diagnosed} prevalences $\pidelta_G,\pidelta_N$ and the corresponding numbers of people diagnosed $\mu_{DG},\mu_{DN}$.  Direct data on any of these parameters would be difficult to obtain.   However, if we were willing to make the assumption in~(\ref{eq:gumcad:diag}), the estimates of diagnosed prevalence would become more precise, for example the posterior median (SD) of $\pidelta_G$ would change from
0.06 (0.13) 
to
0.051 (0.001), though the extent of uncertainty around $\pidelta_N,\mu_{DN}$ would not change substantively.

For the \emph{undiagnosed} prevalences $\pinodelta_G$, $\pinodelta_N$ and undiagnosed case count $\mu_{UG}$, Figure~\ref{fig:res:evppi} shows that more GUM-Anon data (via $\pi^{(GA)}$), more GMSHS data (via $or^{(GM)}$) and more NATSAL data (via $\rho_{UG}$) respectively would give the greatest uncertainty reductions.  These outcomes, however, are already precisely estimated in absolute terms (Figure~\ref{fig:res:prev}).    The number of NGMSM $\mu_{UN}$ with undiagnosed HIV is more uncertain, with 95\% CI (277,1446), and more GMSHS data would be potentially valuable to reduce this uncertainty.

If $\pinodelta_G$ were informed only from the \iftoggle{nodata}{}{4 infections out of }85 people observed in GUM Anon (alternative assumption (a)), the estimates of undiagnosed prevalence or case counts become extremely uncertain, for example, $\var(\mu_{UN})$ increases from $304^2$ to $2859^2$.  We could reduce this uncertainty by collecting more GUM Anon data --- since $EVPPI_{\mu_{UN}}(\pi^{(GA)})$ is $p=62\%$ of $\var(\mu_{UN})$, more GUM Anon data could reduce $\var(\mu_{UN})$ to a minimum of $2859^2(1 - p) = 1770^2$  (note that the square root of the expected variance after learning data is not the same as the expected standard deviation).

\subsection{Partial perfect information for multiple outputs}

Staying with alternative assumption (a), suppose we wish to calculate the maximum potential value of extra GUM Anon data for \emph{jointly} reducing the uncertainty about the number of GMSM, NGMSM and PMSM with undiagnosed HIV, so that $\bm\alpha$ is the vector $(\mu_{UG},\mu_{UN},\mu_{UP})$.  As described in Section~\ref{sec:voi:specific}, we could simply calculate the standard EVPPI based on a scalar output $\alpha$ redefined as their sum, $\mu_U = \mu_{UG}+\mu_{UN}+\mu_{UP}$, the total number of MSM with undiagnosed HIV, whose posterior median is 5164 (SD 3271).  This would ensure that any data expected to reduce the variance of any of these three outputs by the same (additive) amount would be valued equally.  From this, we find that extra GUM Anon data would be expected to reduce $\var(\mu_U)$ from $3271^2$ to a minimum of $1803^2$.    Since $\mu_U$ is dominated by NGMSM (posterior median of $\mu_{UN}$ is 4190), this is mostly explained by an expected reduction in $\var(\mu_{UN})$ from $2859^2$ to a minimum of $1770^2$.

Alternatively, suppose both the prevalences and the case counts are of interest, for example in NGMSM, so that $\bm\alpha = (\pinodelta_N, \mu_{UN})$.  Since these two components are on very different scales, the Bayesian ``D-optimality'' criterion $v(\bm\alpha) = \det(\cov(\bm\alpha))$ would be a preferable measure of overall expected loss due to uncertainty.   We use this criterion to compare the maximum expected value of extra GUM Anon data and extra GMSHS data, which combine to estimate the outcomes for NGMSM as described in Section~\ref{sec:gmshs}.  The EVPPI is interpreted as the expected reduction in the product of $\var(\pinodelta_N)$ and $\var(\mu_{UN})$ given by extra GUM Anon or GMSHS data, adjusted for their covariance. This is 421 and 132 respectively, favouring extra data from GUM Anon.  Though in this example, examining expected reductions in $\var(\pinodelta_N)$ or $\var(\mu_{UN})$ separately would lead to the same conclusion, since $\pinodelta_N$ is defined as the proportion $\mu_{UN} / r_N$ of NGMSM with HIV, and GUM Anon and GMSHS are not informative about the number $r_N$ of NGMSM, thus extra data informs $\mu_{UN}$ entirely through information on $\pinodelta_N$ (or vice versa).

\subsection{Sample information}
\label{res:evsi}

We now estimate the expected value of data with specific sample sizes for improving the precision of the estimated number of people $\mu_U$ with undiagnosed HIV.  Using the GUMCAD data and associated strong prior assumptions, the posterior median of $\mu_U$ is 804 (SD 323), compared to 5164 (SD 3271) with this information excluded.   We compare the value of additional data from GUM Anon and additional data from GMSHS (on top of their original sample sizes of 85 and 945 respectively) for reducing these posterior standard deviations. 

The expected value of sample information (EVSI) is computed for a series of sample sizes $n$ using the method in Section~\ref{sec:evsi:comp}.  For GUM Anon (Section~\ref{sec:gumcad}), the sufficient statistic $T(\y)$ consists of the empirical HIV prevalence $\y/n$ from an additional survey $\y \sim Bin(n, \pi^{(GA)})$.  For GMSHS (Section~\ref{sec:gmshs}), given a sample size $n$, $\y = (N^{(GM)}_G,Y^{(GM)}_G,Y^{(GM)}_N)$, where $N^{(GM)}_G$ is the number of previously-undiagnosed MSM in the future sample of $n$ who attend GUM clinics (the equivalent of the observed $n^{(GM)}_G=493$).  Then $Y^{(GM)}_G$ and $Y^{(GM)}_N$ are the numbers of men out of denominators $N^{(GM)}_G$ and $N^{(GM)}_N=n-N^{(GM)}_G$ (GMSM and NGMSM respectively) who test positive for HIV, the equivalents of the observed $y^{(GM)}_G=20,y^{(GM)}_N=492$. We take $T(\y) = o(\hat p^{(GM)}_N(\y)) / o(\hat p^{(GM)}_G(\y))$, a point estimator of the odds ratio, where $\hat p^{(GM)}_G(\y)$ is an estimator of the proportion of MSM in group $g$ who have HIV.  To avoid zeros in the denominator $o(\hat p^{(GM)}_G(\y))$, we use a Bayesian estimator $\hat p^{(GM)}_G(\y) = (Y^{(GM)}_G+0.5)/(N^{(GM)}_G+1)$, the posterior mean of a binomial proportion under a Jeffreys Beta(0.5,0.5) prior, rather than the empirical proportion $Y^{(GM)}_G/N^{(GM)}_G$.

\begin{figure}
\begin{knitrout}
\definecolor{shadecolor}{rgb}{0.969, 0.969, 0.969}\color{fgcolor}
\includegraphics[width=\maxwidth]{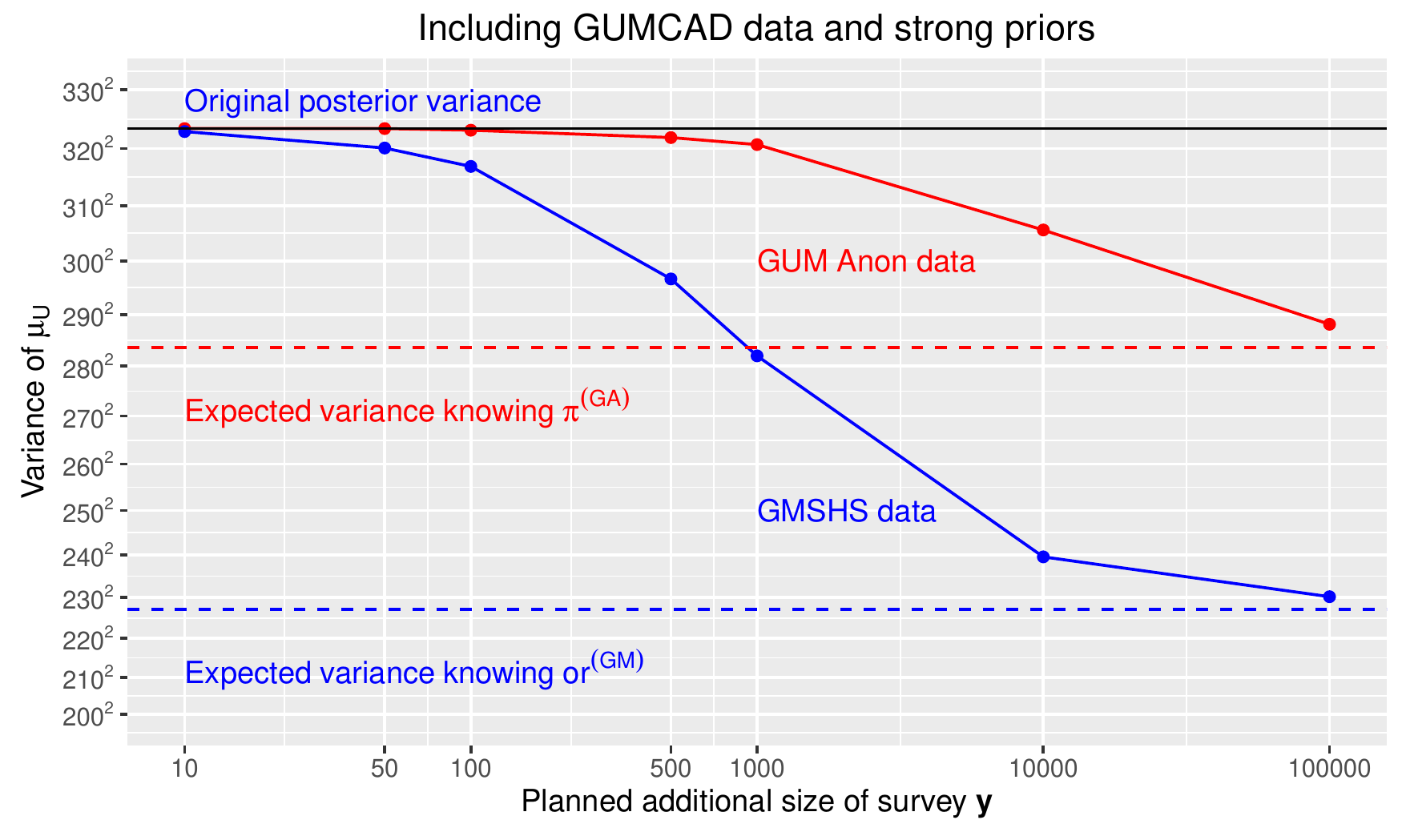} 

\includegraphics[width=\maxwidth]{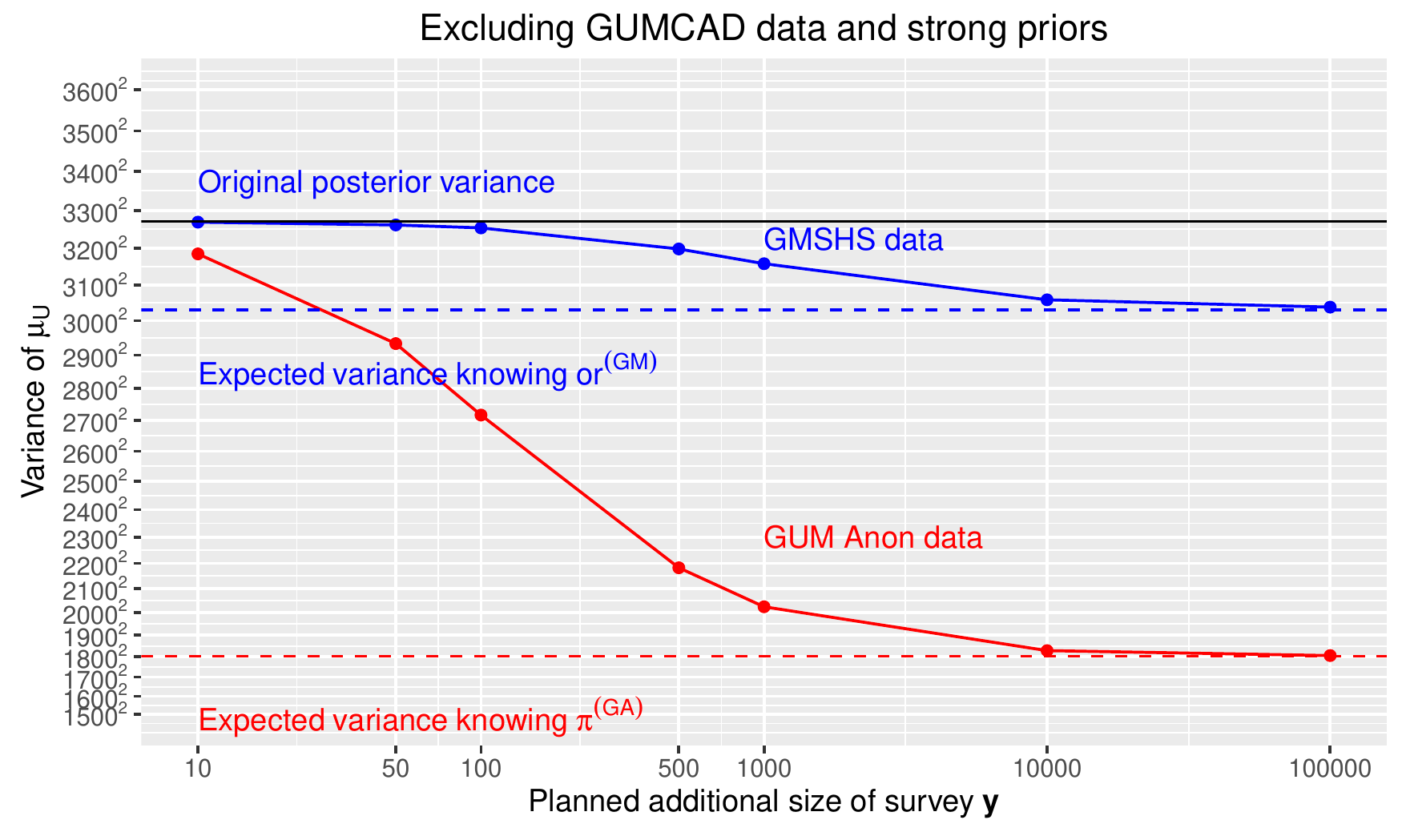} 

\end{knitrout}
  \caption{Expected value of sample information: value of additional data from GUM Anon or GMSHS for reducing the variance of the total number of MSM with undiagnosed HIV, $\mu_U = \mu_{UG}+\mu_{UN}$.  The $x$ axis is on the log scale. The $y$ axis is the variance, with the labels as SD$^2$. }
  \label{fig:res:evsi}
\end{figure}

Figure~\ref{fig:res:evsi} shows $\var(\mu_U) - EVSI(\y)$, the expected variance remaining after data collection, under the two alternative assumptions.     With the strong priors, $\mu_U$ is relatively well informed, and extra data from GUM Anon at realistic sample sizes (1000 or less) would not noticeably reduce $\var(\mu_U)$.   GMSHS data would be more valuable, through improving the estimate of $\mu_{UN}$, the more uncertain contributor to $\mu_U = \mu_{UG} + \mu_{UN}$.   1000 extra observations from GMSHS would be expected to reduce $\var(\mu_U)$ from $323^2$ to $282^2$.

Without the strong prior information, $\var(\mu_U)=3271^2$ is substantially greater, and $\mu_U$ is only directly informed by the 85 observations from GUM Anon.  Extra data from this source would be valuable, for example, another 500 observations would be expected to reduce this variance to $2183^2$.   Relative to these improvements, GMSHS data of the same size would be much less valuable.   GMSHS data however would be expected to give around the same \emph{absolute} reductions in $\var(\mu_U)$, whether or not the strong priors are included.

\section{Summary and potential further work}

\label{sec:conc}

We have presented tools to find the most influential sources of uncertainty in a multiparameter evidence synthesis context and determine the expected value of extra data.  We generalized methods, previously only applied in deterministic models, to complex graphical models, a class which also includes hierarchical models.   We have shown how VoI methods developed for formal finite-choice decision problems can be extended to deal with estimation of single or multiple quantities.   Therefore the same methods can be used for formal decision problems based on graphical models, e.g. an HIV prevalence estimation model such as ours could be used to compare strategies for HIV testing.   This would allow the optimal sample size of future data to be determined, through a health economic loss that trades off the cost of data collection with the expected health benefits gained from extra information that reduces the probability of choosing a sub-optimal policy.

In the HIV application, we found that structural assumptions, such as whether to include a particular piece of information, were influential to both the parameter estimates and the value of information.   Such uncertainties might be parameterised \citep[see, e.g.][]{strong:discrepancy:jrssc}, for example a particular prior or dataset of uncertain relevance could be discounted using an unknown weight \citep[e.g.][]{neuenschwander2009note}.  The EVPPI of the extra parameter would then quantify this uncertainty in the context of all other uncertainties, referred to as the ``expected value of model improvement'' by \citet{strong2014model}. 

Note that VoI refers to the expected value of \emph{potential future} information, which differs from the \emph{observed} value of a dataset $x_i$ \emph{currently} included in the model.   The latter could be computed as the observed reduction in loss when the model is refitted without $x_i$.  This could demonstrate the value of past data to the policymaker responsible for funding the collection of future data of the same type.   For surveys or longitudinal studies conducted at regular intervals, VoI might be used to determine the expected value of future surveys or follow-up, although a full analysis would require modelling the expected changes through time in the quantities, such as disease prevalence or incidence, informed by the data. 

While our method is broadly applicable, the details of computation for different decision problems and loss functions will be different.  We discussed finite-action decisions and point estimation.  A more general decision problem is to  estimate the entire uncertainty distribution of $\bm\theta$.  The standard posterior $p(\bm\theta|\y)$ is then optimal under a log scoring rule \citep{bernardo:smith}, and \citep[following][]{lindley1956measure} standard Bayesian design theory aims to maximise the information gain from new data $\y$, which we can write as $EVSI(\y) =  E_{\bm\theta}(-\log(p(\bm\theta))) + E_{\y} E_{\theta|\y} \{\log(p(\bm\theta|\y))$.  Under linear models~\citep{chaloner1995bayesian}, this is equivalent to minimising $\det(\cov(\bm\theta))$, but more generally this is challenging to compute \citep{ryan2015review}.

Note that the VoI approach to sensitivity analysis is an example of the ``global'' approach, which examines the changes in model outputs given by varying parameters within the ranges of their belief distributions.  The ``local'' approach is based on examining the posterior geometry resulting from small parameter perturbations around a base case, e.g. \citet{roos2015sensitivity} assess the robustness of hierarchical models to prior assumptions in this way.  While the global approach is easier to interpret, as discussed by \citet{oakley:ohagan:psa} and \citet{roos2015sensitivity}, it conditions on one particular prior specification, and parameterising all potential prior beliefs or structural assumptions would be impractical. 

The regression method for VoI computation that we described requires only a MCMC sample from the joint distribution of parameters of interest $\bm\phi$ and outputs $\bm\alpha$.  Additionally for EVSI it requires that the information in the new data $\y$ can be condensed into an analytic sufficient statistic $T(\y)$.   Alternative methods which exploit particular analytic structures of $g()$, where $\alpha$ is a known function $g(\bm\phi)$, thus avoiding a regression approximation, were discussed by \citet{madan2014strategies} for EVPPI and and \citet{ades2004expected} for EVSI.    \citet{menzies2016efficient} also presented an importance resampling method for EVSI computation which needs only a single MCMC sample and not a sufficient statistic. 

In conclusion, the consideration of future evidence requirements is an often-neglected part of statistical analysis.  The Value of Information methods we have presented provide a practicable set of tools for achieving this aim in the context of Bayesian evidence synthesis.

\appendix
\section*{Appendix: Supplementary figures}

\begin{figure}
\begin{knitrout}
\definecolor{shadecolor}{rgb}{0.969, 0.969, 0.969}\color{fgcolor}
\includegraphics[width=\maxwidth]{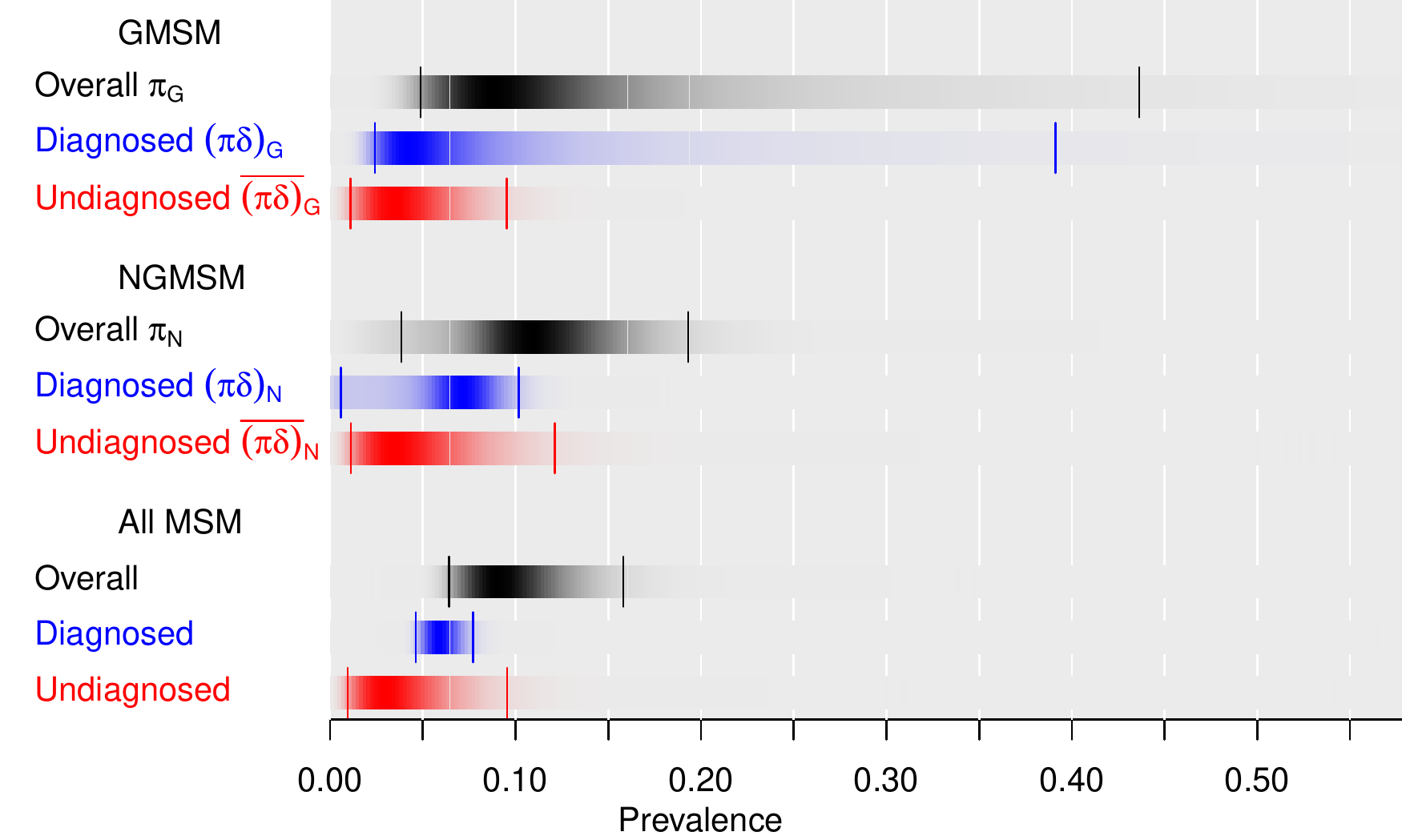} 

\includegraphics[width=\maxwidth]{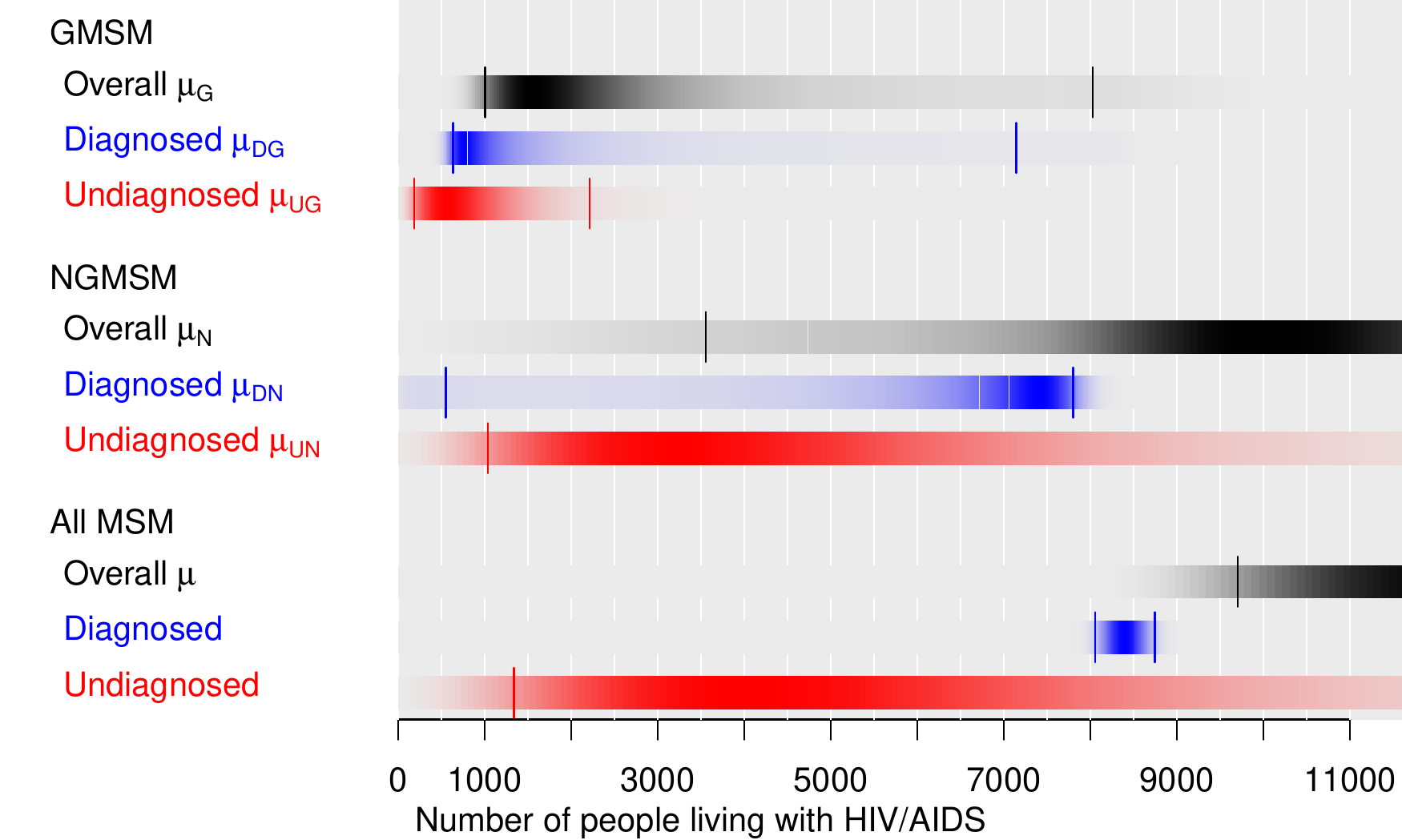} 

\end{knitrout}
  \caption{Posterior distributions of HIV prevalence (top) and numbers of MSM living with HIV/AIDS (bottom), London 2012. Darkness within each strip proportional to posterior density, with 95\% credible intervals indicated. Alternative scenario (a): undiagnosed prevalence from GUM Anon only }
  \label{fig:res:prev}
\end{figure}

\begin{figure}
\begin{knitrout}
\definecolor{shadecolor}{rgb}{0.969, 0.969, 0.969}\color{fgcolor}
\includegraphics[width=\maxwidth]{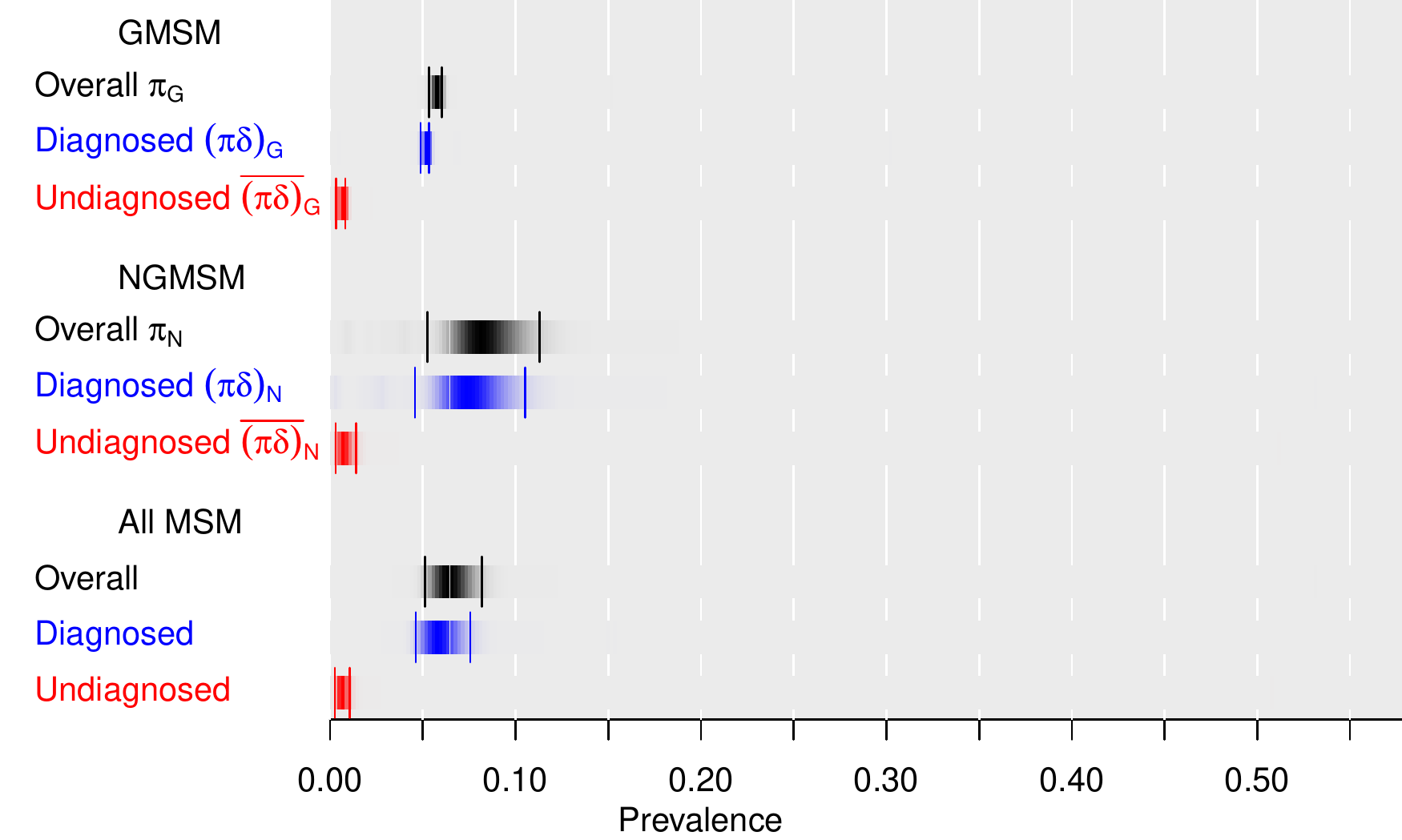} 

\includegraphics[width=\maxwidth]{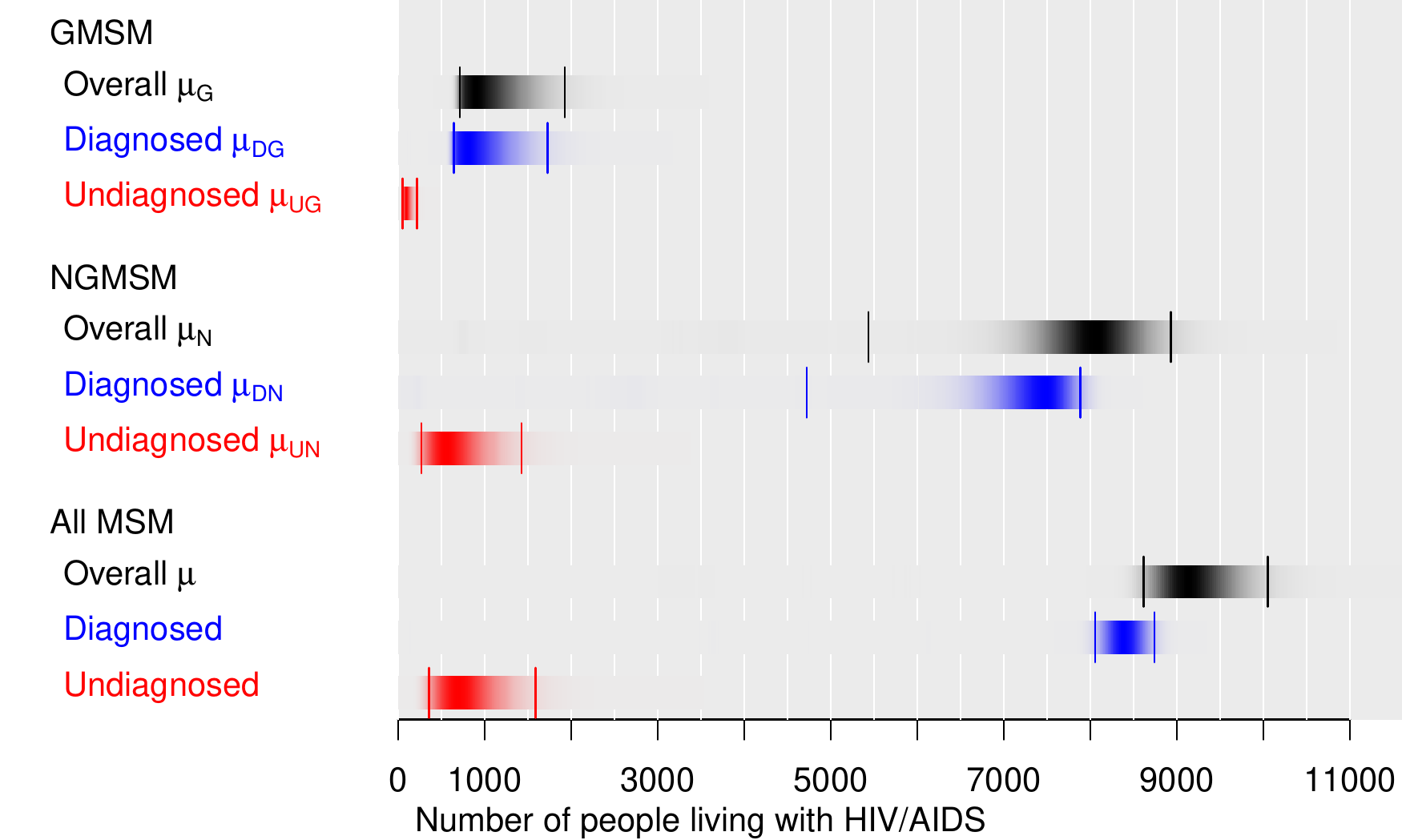} 

\end{knitrout}
  \caption{Posterior distributions of HIV prevalence (top) and numbers of MSM living with HIV/AIDS (bottom), London 2012. Darkness within each strip proportional to posterior density, with 95\% credible intervals indicated. Alternative scenario (b): GUMCAD also informs diagnosed prevalence }
  \label{fig:res:prev}
\end{figure}

\begin{figure}
\begin{knitrout}
\definecolor{shadecolor}{rgb}{0.969, 0.969, 0.969}\color{fgcolor}
\includegraphics[width=\maxwidth]{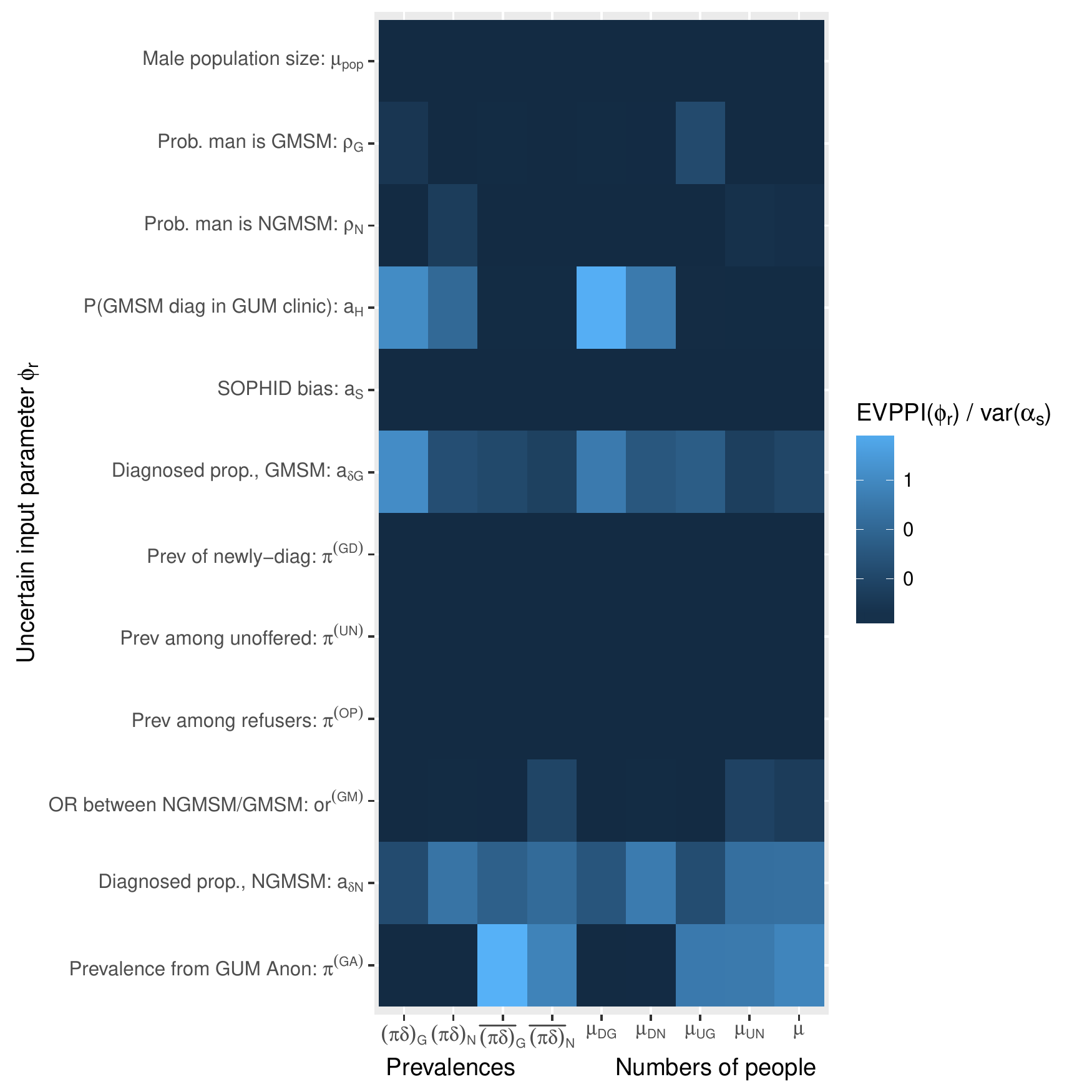} 

\end{knitrout}
  \caption{Expected value of partial perfect information in the HIV prevalence model.  Alternative scenario (a): undiagnosed prevalence from GUM Anon only }
  \label{fig:res:evppi}
\end{figure}

\begin{figure}
\begin{knitrout}
\definecolor{shadecolor}{rgb}{0.969, 0.969, 0.969}\color{fgcolor}
\includegraphics[width=\maxwidth]{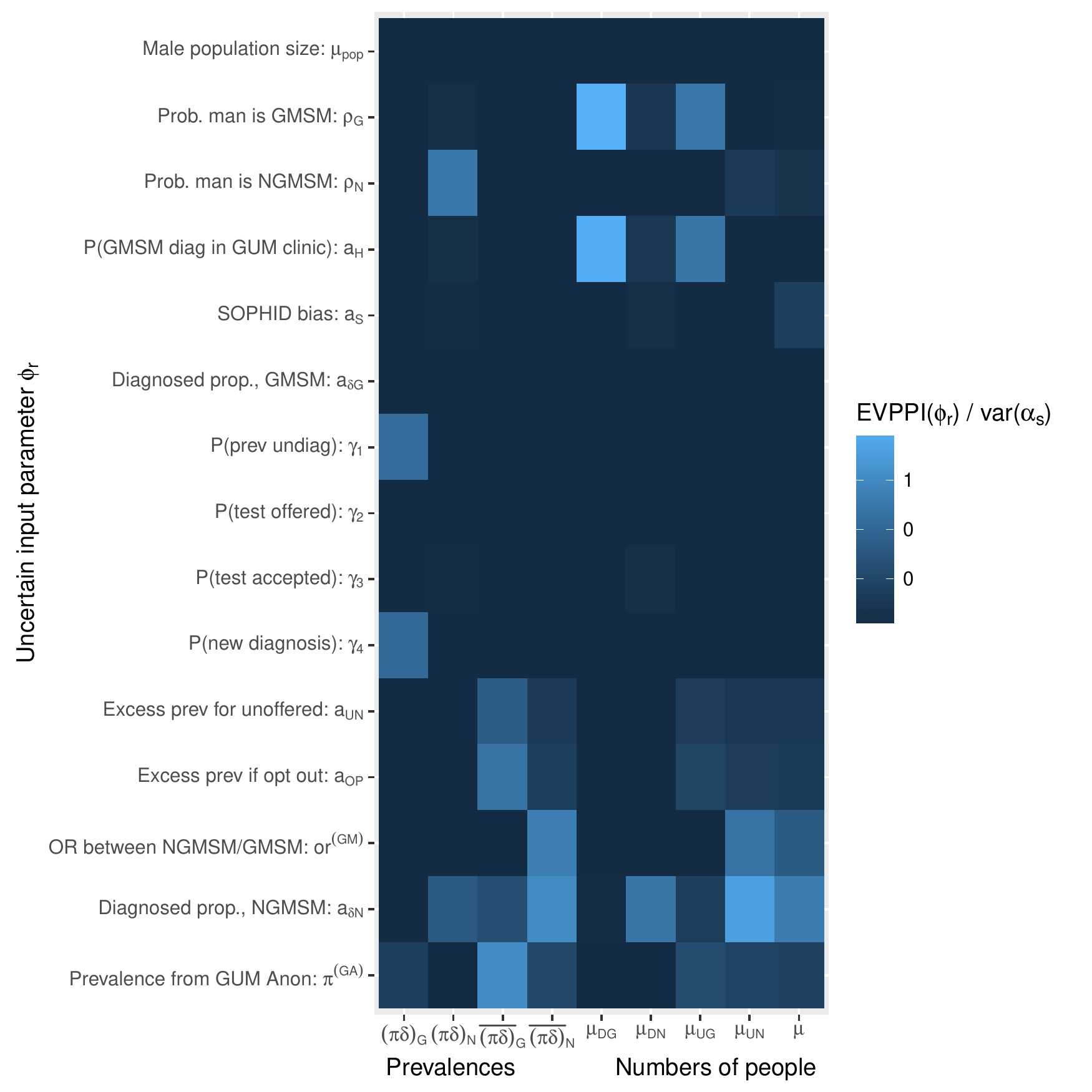} 

\end{knitrout}
  \caption{Expected value of partial perfect information in the HIV prevalence model.  Base case analysis in paper. Alternative scenario (b): GUMCAD also informs diagnosed prevalence }
  \label{fig:res:evppi}
\end{figure}

%% http://tex.stackexchange.com/questions/41821/creating-bib-file-containing-only-the-cited-references-of-a-bigger-bib-file
%\bibliography{abbrev,bayesian,uncertainty,evi,voibayes,stats,cost,comp}

\end{document}